\documentclass{aa}  
\usepackage{verbatim}
\usepackage{graphicx}

\usepackage{txfonts}
\usepackage{tikz}
\usetikzlibrary{arrows.meta, positioning, quotes}

\usepackage[colorlinks=true,citecolor=blue]{hyperref}

\newcommand{\fixme}[1]{{#1}}
\newcommand{\revision}[1]{{#1}}

\begin{document} 
   \title{Galaxy infall models for arbitrary velocity directions}

   \author{Jenny Wagner\inst{1,2,3}
          \and David Benisty\inst{4}   }

   \institute{Helsinki Institute of Physics, P.O. Box 64, FI-00014 University of Helsinki, Finland
        \and 
        Academia Sinica Institute of Astronomy and Astrophysics, 11F of AS/NTU Astronomy-Mathematics Building, No.1, Sec.~4, Roosevelt Rd, Taipei 106216, Taiwan, R.O.C\\
             \email{wagner@asiaa.sinica.edu.tw}
        \and 
        Bahamas Advanced Study Institute and Conferences, 4A Ocean Heights, Hill View Circle, Stella Maris, Long Island, The Bahamas
        \and Leibniz-Institut f\"ur Astrophysik Potsdam (AIP), An der Sternwarte 16, 14482 Potsdam, Germany\\
              \email{benidav@aip.de}  }

   \date{Received XXX; accepted YYY}

  \abstract
{ \fixme{For most galaxies in the cosmos, our knowledge of their motion is limited to line-of-sight velocities from redshift observations. To determine the radial velocity between two galaxies the minor and major infall models were established by Karachentsev \& Kashibadze
(2006). 
Regardless of the background cosmology, our derivations reveal that these infall models approximate the total radial velocity between two galaxies by two different projections employing different information about the system. 
For galaxies having small angular separations $\theta$, all infall models agree that the radial velocity is the difference of their line-of-sight components. 
Applying these models to ca.~$500$ halos of the Illustris-3 simulation, we find the perpendicular and tangential velocity parts to be non-negligible for more than 90\% of all, more than 5000 infalling subhalos. 
Thus, even for $\theta < 10$~deg, the infall-model velocities deviate from the true radial velocity. 
Only for 30\% we found the true one lay between the minor and major infall velocity. However, the infall models yield robust upper and lower bounds to the true radial velocity dispersion. 
Observed under $\theta < 10$~deg the velocity dispersion inferred from the sole difference of line-of-sight velocity components even coincides with the true one, justifying this approach for high-redshift groups and clusters. 
Based on these findings, we predict the radial velocity dispersion of the M81-group from the minor infall model (upper bound) $\sigma_{\mathrm{r,min}} = (\revision{180} \pm 42)~\mbox{km}/\mbox{s}$, from the major infall model (lower bound) $\sigma_{\mathrm{r,maj}} = (\revision{142} \pm 64) ~\mbox{km}/\mbox{s}$ and $\sigma_\mathrm{r,\Delta v} = (\revision{99} \pm 36)~\mbox{km}/\mbox{s}$ from the line-of-sight-velocity difference.}}

\maketitle

\nolinenumbers  
\section{Introduction}
\label{sec:introduction}

Studying the relative motion between cosmic structures is complicated for observers like us who are located at a random external position.
In the majority of cases, external observers only measure velocity components along their lines of sight to the cosmic objects. 
This is often done by highly precise spectroscopic redshift observations. 
Most objects, like galaxies, do not have a directly observable peculiar motion on the observer's celestial sky. 
For instance, the peculiar motion for our neighbour-galaxy, M31, was inferred from peculiar-motion measurements of its stars and modelling its satellite dwarf galaxy motions \citep{bib:Sohn2012, bib:Marel2012}.

Depending on the distance to us, spectroscopic redshifts need to be partitioned into a contribution from a cosmological background model and the velocity on top of it. 
The latter task is difficult in our cosmic neighbourhood, where both contributions play an equally important role, see recent observations and estimates by Cosmicflows \citep{bib:Tully2023, bib:Valade2024} or the Dark Energy Spectroscopic Instrument \citep{bib:DESI2024,bib:Said2024}.

Reconstructing a self-consistent map of our cosmic neighbourhood of mass (density) distributions from galaxies, groups to clusters and their motions, many models have been used.
Two seminal works were \cite{bib:Karachentsev2006} and \cite{bib:Karachentsev2010}.
They considered two limiting cases: the minor infall model assumes that galaxies are mainly moved by an expanding cosmological background, a linear Hubble flow, and are only subject to a small mutual gravitational attraction. The major infall model assumes that galaxies fall into the gravitational potential of a galaxy group or cluster instead of following the Hubble flow.

Both models neglect the velocity components perpendicular to the observer's line of sight, they assume that the structure is spherically symmetric to calculate the radial infall velocity and that it is embedded in a linear Hubble flow.
The authors then tried to select applications that obey these assumptions well and reduce the impact of the unknown tangential velocity component at the same time.
\cite{bib:Karachentsev2006} applied both models to the Local Group and the M81/M82-group, \cite{bib:Karachentsev2010} analysed the Virgo cluster.
In \citep{bib:Karachentsev2006} they probed the impact of the tangential velocity by a simulation. 
Following-up, \cite{bib:Kim2020} analysed the Hubble flow around Virgo and used minor and major infall velocities, after \cite{bib:Sorce2016} only used the minor infall for a similar study.
\fixme{Even though there are only a few papers mentioning the names of these infall models, many works use the minor infall model to estimate the radial velocity component of binaries or galaxy groups, as, for instance, \cite{bib:Diaz2014,bib:Penarrubia2014}.}

To overcome these restrictions and to generalise the models for a deeper analytical understanding, we describe a general binary motion without assuming spherical symmetry or specifying the embedding cosmology.
Staying on this general, kinematics-only level, it remains open if the motions occur in a gravitationally bound or unbound volume. 
We also include the perpendicular velocity components into the models, as detailed in Sect.~\ref{sec:pairwise_infall}. 
For the first time, we quantify their impact on the infall models and derive conditions under which the infall models yield lower or upper bounds to the true radial infall velocity. 
Analogously, we investigate the role of the tangential velocity.
We also compare the two infall models with each other to derive the general conditions under which both models coincide.  
In Sect.~\ref{sec:galaxy_clusters}, we then extend the formalism to galaxy groups or clusters and investigate the impact of the statistics on the accuracy of the models for larger structures.
\fixme{Section~\ref{sec:simulation} applies the infall models to cosmic structures from a snapshot at redshift $z=0$ simulated in the Illustris-3 dataset that is publicly available online\footnote{\url{https://www.tng-project.org/data/docs/scripts/}} in order to investigate the limits of the models in a realistic setting. 
Subsequently, in Sect.~\ref{sec:application}, we revisit the M81-system to discuss the accuracy of the reconstruction that is possible given the findings in this work.}
Finally, Sect.~\ref{sec:conclusions} summarises our results and gives an outlook on the applicability of the generalised models.

\section{Pairwise infall models}
\label{sec:pairwise_infall}

\subsection{Definitions and notation}
\label{sec:definitions}

To simplify calculations, we use three-dimensional vectors in bold-font, like $\boldsymbol{r}$.
Their amplitudes\footnote{Amplitudes $\left| \cdot \right|$ can be negative, while the absolute (non-negative) size of a vector is denoted as $\lVert \cdot \rVert_2$.} are denoted with $\left| \boldsymbol{r} \right|$ and their directions in terms of unit vectors by $\hat{\boldsymbol{r}}$. 
Line-of-sight components are denoted by subscript l, while components perpendicular to the line of sight are denoted with $\perp$ as subscript. 
By construction, their scalar product vanishes, $\boldsymbol{r}_{\mathrm{l}} \cdot \boldsymbol{r}_{\perp} = 0$.

\begin{figure}
\label{fig:general_model}
\centering
\includegraphics[width=0.4\textwidth]{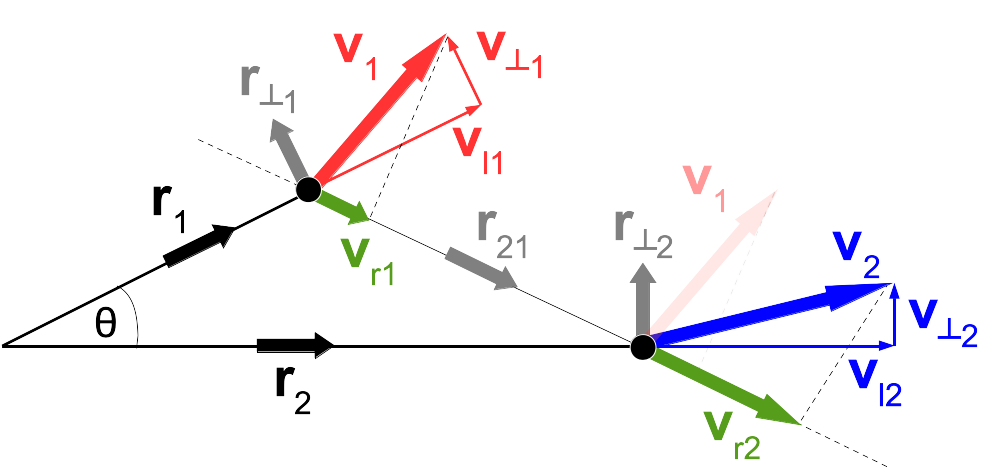}
\caption{Three-dimensional motion of two galaxies and definition of notations. Maximum information measurable for a distant observer are velocity components along the line of sight $\boldsymbol{v}_{\mathrm{l} 1}$, $\boldsymbol{v}_{\mathrm{l} 2}$ to high precision via spectroscopy, highly precise angle between the galaxy positions on the sky, $\theta$, and line-of-sight distances $\boldsymbol{r}_1, \boldsymbol{r}_2$ with a precision depending on the probe used, see, \cite{bib:Tully2023} for recent examples.}
\end{figure}

In this notation, let galaxy $i$ at redshift $z_i$ be at a (cosmological) distance $\boldsymbol{r}_i$, moving at a total velocity $\boldsymbol{v}_i$ with respect to this observer. 
We partition $\boldsymbol{v}_i$ into a projection along the observer's line of sight and one perpendicular to it as
\begin{equation}
\boldsymbol{v}_{\mathrm{l}i} \equiv \left( \boldsymbol{v}_i \cdot \hat{\boldsymbol{r}}_i \right) \, \hat{\boldsymbol{r}}_i = \left| \boldsymbol{v}_{\mathrm{l} i} \right| \, \hat{\boldsymbol{r}}_i \;, \quad \boldsymbol{v}_{\perp i} \equiv \left( \boldsymbol{v}_i \cdot \hat{\boldsymbol{r}}_{\perp i} \right) \,  \hat{\boldsymbol{r}}_{\perp i} = \left| \boldsymbol{v}_{\perp i} \right| \,  \hat{\boldsymbol{r}}_{\perp i} \;.
\label{eq:definitions}
\end{equation} 
Considering arbitrary motions of two galaxies $i=1,2$ with distances from the observer $\boldsymbol{r}_1$ and $\boldsymbol{r}_2$ and total velocities $\boldsymbol{v}_1$ and $\boldsymbol{v}_2$, respectively, Fig.~\ref{fig:general_model} shows all distances, and velocity components.
To split every vector in its amplitude and direction, we use $\left| \boldsymbol{r}_i \right| \equiv \lVert \boldsymbol{r}_i \rVert_2$, $i=1,2$ and read off Fig.~\ref{fig:general_model} that
\begin{eqnarray}
\left| \boldsymbol{r}_{21} \right| &\equiv& \Big( \lVert \boldsymbol{r}_1\rVert_{2}^2 + \lVert \boldsymbol{r}_2\rVert_{2}^2 - 2\lVert \boldsymbol{r}_{1}\rVert_2 \lVert \boldsymbol{r}_{2}\rVert_2 \cos\theta \Big)^{1/2} \label{eq:ur1} \;, \\
\hat{\boldsymbol{r}}_{21} &=& \frac{\boldsymbol{r}_{2} - \boldsymbol{r}_{1}}{\left| \boldsymbol{r}_{21}\right|} = \frac{\left| \boldsymbol{r}_2 \right| \hat{\boldsymbol{r}}_{2} - \left| \boldsymbol{r}_1 \right| \hat{\boldsymbol{r}}_{1}}{\left| \boldsymbol{r}_{21} \right|} \equiv \overline{r}_2 \hat{\boldsymbol{r}}_{2} - \overline{r}_1 \hat{\boldsymbol{r}}_{1} \label{eq:ur2} \;, \\
\hat{\boldsymbol{r}}_{21} &=& \left( \overline{r}_2 \cos\theta - \overline{r}_1 \right) \hat{\boldsymbol{r}}_1 - \overline{r}_2 \sin \theta \,\hat{\boldsymbol{r}}_{\perp 1} \;, \label{eq:ur7} \\
&=&  \left( \overline{r}_2 - \overline{r}_1 \cos \theta \right) \hat{\boldsymbol{r}}_2 - \overline{r}_1 \sin \theta \, \hat{\boldsymbol{r}}_{\perp 2} \;, \label{eq:ur8} \\
\hat{\boldsymbol{r}}_i \cdot \hat{\boldsymbol{r}}_{\perp i} &=& 0 \;, \quad i =1,2 \label{eq:ur3} \;, \\
\hat{\boldsymbol{r}}_1 \cdot \hat{\boldsymbol{r}}_2 &=& \hat{\boldsymbol{r}}_{\perp 1} \cdot \hat{\boldsymbol{r}}_{\perp 2} = \cos\theta  \label{eq:ur4} \;, \\
\hat{\boldsymbol{r}}_1 \cdot \hat{\boldsymbol{r}}_{\perp 2} &=& - \hat{\boldsymbol{r}}_2 \cdot \hat{\boldsymbol{r}}_{\perp 1} = \sin\theta \label{eq:ur5} \;. 
\end{eqnarray}
Eq.~\eqref{eq:ur5} can acquire a minus depending on the definition of $\boldsymbol{r}_{\perp i}$. Here, it follows the definition of Fig.~\ref{fig:general_model}.
The directions of $\boldsymbol{r}_i$ are chosen \fixme{to be positive from the observer towards the galaxies. 
Hence the velocity amplitudes $\left| \boldsymbol{v}_{\mathrm{l} i} \right|$ can be positive or negative for $i=1,2$, depending on the relative orientation of the velocity with respect to the distance vector}. 
There is no unique $\boldsymbol{r}_{\perp i}$ to $\boldsymbol{r}_i$ in three dimensions.
Since only $\boldsymbol{r}_i$ and projections onto this vector yield observables, the non-uniqueness of $\boldsymbol{r}_{\perp i}$ implies that this component does not have a known direction in space.
Only parts can be constrained by a projection onto the observer's sky.
Due to this freedom, $\boldsymbol{r}_{\perp i}$ are defined such that $\left| \boldsymbol{v}_{\perp i} \right| \ge 0$.

Similarly, we can express the motion perpendicular to $\hat{\boldsymbol{r}}_{21}$.
To calculate $\hat{\boldsymbol{r}}_{\perp 21}$, we define
\begin{eqnarray}
\hat{\boldsymbol{n}}_{1} &=& \overline{r}_2 \sin \theta \, \hat{\boldsymbol{r}}_1 + \left(\overline{r}_2 \cos \theta - \overline{r}_1 \right)  \hat{\boldsymbol{r}}_{\perp 1} \;, \\
\hat{\boldsymbol{n}}_{2} &=& \overline{r}_1 \sin \theta \, \hat{\boldsymbol{r}}_2 + \left(\overline{r}_2 - \overline{r}_1 \cos \theta \right)  \hat{\boldsymbol{r}}_{\perp 2} \;,
\end{eqnarray}
which obey $\left| \hat{\boldsymbol{n}}_i \right| = 1$, $\hat{\boldsymbol{r}}_{21} \cdot \hat{\boldsymbol{n}}_{i} = 0$, and $\hat{\boldsymbol{n}}_{1} \cdot \hat{\boldsymbol{n}}_{2} = 1$, $i=1,2$.
Since $\boldsymbol{v}_i = \boldsymbol{v}_{\mathrm{l} i} + \boldsymbol{v}_{\perp i}$ and there is no velocity component simultaneously perpendicular to $\hat{\boldsymbol{r}}_i$ and $\hat{\boldsymbol{r}}_{\perp i}$,  $\hat{\boldsymbol{r}}_{\perp 21} \equiv \hat{\boldsymbol{n}}_i$, $i=1,2$.

Based on these relations, we derive
\begin{eqnarray}
\hat{\boldsymbol{r}}_1 \cdot \hat{\boldsymbol{r}}_{21}    &=& \overline{r}_2 \cos \theta - \overline{r}_1 \label{eq:ur9} \;, \\
  \hat{\boldsymbol{r}}_2 \cdot \hat{\boldsymbol{r}}_{21}  &=& \overline{r}_2 - \overline{r}_1 \cos \theta \label{eq:ur10} \;, \\
  \hat{\boldsymbol{r}}_{\perp 1} \cdot \hat{\boldsymbol{r}}_{21}  &=& - \overline{r}_2 \sin \theta \label{eq:ur11} \;, \\ 
\hat{\boldsymbol{r}}_{\perp 2} \cdot \hat{\boldsymbol{r}}_{21}  &=& - \overline{r}_1 \sin \theta \label{eq:ur12} \;, \\
\hat{\boldsymbol{r}}_1 \cdot \hat{\boldsymbol{r}}_{\perp 21}    &=& \overline{r}_2 \sin\theta \label{eq:ur13} \;, \\
\hat{\boldsymbol{r}}_2 \cdot \hat{\boldsymbol{r}}_{\perp 21}    &=& \overline{r}_1 \sin\theta \label{eq:ur14} \;, \\
\hat{\boldsymbol{r}}_{\perp 1} \cdot \hat{\boldsymbol{r}}_{\perp 21}    &=& \left(\overline{r}_2 \cos \theta - \overline{r}_1 \right) \label{eq:ur15} \;, \\
\hat{\boldsymbol{r}}_{\perp 2} \cdot \hat{\boldsymbol{r}}_{\perp 21}    &=& \left(\overline{r}_2 - \overline{r}_1 \cos \theta \right) \label{eq:ur16} \;.
\end{eqnarray}

\subsection{General infall model}
\label{sec:general_infall}

In the notation of Sect.~\ref{sec:definitions}, the relative velocity between the two galaxies reads
\begin{align}
\boldsymbol{v}_{21} \equiv  \boldsymbol{v}_2 - \boldsymbol{v}_1 &= \boldsymbol{v}_{\mathrm{l} 2} - \boldsymbol{v}_{\mathrm{l} 1} + \boldsymbol{v}_{\perp 2} - \boldsymbol{v}_{\perp 1} \label{eq:vel_dif} \\
&=  \left| \boldsymbol{v}_{\mathrm{l} 2} \right| \hat{\boldsymbol{r}}_{2} - \left| \boldsymbol{v}_{\mathrm{l} 1} \right| \hat{\boldsymbol{r}}_{1} + \left| \boldsymbol{v}_{\perp 2} \right| \hat{\boldsymbol{r}}_{\perp 2} - \left| \boldsymbol{v}_{\perp 1} \right| \hat{\boldsymbol{r}}_{\perp 1} \;.
\label{eq:vel_diff_proj}
\end{align}
To infer their radial infall velocity, we project Eq.~\eqref{eq:vel_diff_proj} onto their connection line. 
Using Eqs.~\eqref{eq:ur1} to \eqref{eq:ur16}, we obtain
\begin{align}
  \left| \boldsymbol{v}_{\mathrm{r}} \right| &\equiv  \left( \boldsymbol{v}_2 - \boldsymbol{v}_1 \right) \cdot \hat{\boldsymbol{r}}_{21} 
  = \left( \boldsymbol{v}_{\mathrm{l} 2} - \boldsymbol{v}_{\mathrm{l} 1} \right) \cdot \hat{\boldsymbol{r}}_{21}  + \left( \boldsymbol{v}_{\perp 2} - \boldsymbol{v}_{\perp 1} \right) \cdot \hat{\boldsymbol{r}}_{21} \\
&= \left| \boldsymbol{v}_{\mathrm{l} 1} \right| \bar{r}_1 + \left| \boldsymbol{v}_{\mathrm{l} 2} \right| \bar{r}_2 - \cos \theta  \left( \left| \boldsymbol{v}_{\mathrm{l} 1} \right| \bar{r}_2 + \left| \boldsymbol{v}_{\mathrm{l} 2} \right| \bar{r}_1 \right) \label{eq:v_r_proj} \\
& \quad \qquad \qquad \qquad + \sin\theta \left( \left| \boldsymbol{v}_{\perp 1} \right| \bar{r}_2 - \left| \boldsymbol{v}_{\perp 2} \right|\bar{r}_1 \right) \nonumber \;.
\end{align}
Thus, the radial velocity is expressed via the observable quantities $\theta$, $\boldsymbol{r}_1$, $\boldsymbol{r}_2$, $\boldsymbol{v}_{\mathrm{l 1}}$, and $\boldsymbol{v}_{\mathrm{l 2}}$ and the unknown quantities $\boldsymbol{v}_{\perp 1}$ and $\boldsymbol{v}_{\perp 2}$. 
As a derived quantity, $\left| \boldsymbol{v}_{\mathrm{r}} \right|$ can have negative, positive, or zero values.
To linear order in $\theta$, Eq.~\eqref{eq:v_r_proj} reads
\begin{equation}
\left| \boldsymbol{v}_{\mathrm{r}} \right| \approx  \left| \boldsymbol{v}_{\mathrm{l} 2} \right| -  \left| \boldsymbol{v}_{\mathrm{l} 1} \right| + \theta \left( \left| \boldsymbol{v}_{\perp 1} \right|\overline{r}_2 - \left| \boldsymbol{v}_{\perp 2} \right|\overline{r}_1 \right)\;,
\label{eq:v_r_linear}
\end{equation}
such that $\left| \boldsymbol{v}_{\mathrm{r}} \right|$ is the difference between the line-of-sight components plus a perturbation. 

Similarly, projecting Eq.~\eqref{eq:vel_dif} onto $\hat{\boldsymbol{r}}_{\perp 21}$, we obtain the velocity component perpendicular to the radial one
\begin{eqnarray}
\left| \boldsymbol{v}_{\mathrm{t}} \right| &\equiv&  \left( \boldsymbol{v}_2 - \boldsymbol{v}_1 \right) \cdot \hat{\boldsymbol{r}}_{\perp 21} \\
&=& \left( \boldsymbol{v}_{\mathrm{l} 2} - \boldsymbol{v}_{\mathrm{l} 1} \right) \cdot \hat{\boldsymbol{r}}_{\perp 21} + \left( \boldsymbol{v}_{\perp 2} - \boldsymbol{v}_{\perp 1} \right) \cdot \hat{\boldsymbol{r}}_{\perp 21} \\
&=& \sin \theta \left( \left| \boldsymbol{v}_{\mathrm{l} 2} \right| \overline{r}_1 - \left| \boldsymbol{v}_{\mathrm{l} 1} \right| \overline{r}_2 \right) \label{eq:v_t_proj} \\
& & + \left| \boldsymbol{v}_{\perp 2} \right| \overline{r}_2 + \left| \boldsymbol{v}_{\perp 1} \right| \overline{r}_1 - \cos \theta \left(\left| \boldsymbol{v}_{\perp 1} \right| \overline{r}_2 + \left| \boldsymbol{v}_{\perp 2} \right| \overline{r}_1 \right)  \nonumber \\
&\approx& \left| \boldsymbol{v}_{\perp 2} \right| - \left| \boldsymbol{v}_{\perp 1} \right| + \theta \left( \left| \boldsymbol{v}_{\mathrm{l} 2} \right| \overline{r}_1 - \left| \boldsymbol{v}_{\mathrm{l} 1} \right| \overline{r}_2 \right) \;.
\label{eq:v_t_linear}
\end{eqnarray} 
We denote this component the tangential velocity.
As a derived quantity, it can have any negative, positive, or zero value.
To leading order, $\left| \boldsymbol{v}_{\mathrm{t}} \right|$ is proportional to the difference of the perpendicular velocity components plus a perturbation.

We can also express the velocity difference as a combination of radial and tangential components
\begin{equation}
\boldsymbol{v}_2 - \boldsymbol{v}_1 \equiv \boldsymbol{v}_{\mathrm{r}} + \boldsymbol{v}_{\mathrm{t}} = \left| \boldsymbol{v}_{\mathrm{r}}\right| \hat{\boldsymbol{r}}_{21} + \left| \boldsymbol{v}_{\mathrm{t}} \right| \hat{\boldsymbol{r}}_{\perp 21} \;.
\label{eq:v_rad_tan}
\end{equation}

Since $\boldsymbol{v}_{\mathrm{t}}$ is not observable in general, we read off Eq.~\eqref{eq:v_t_proj}, under which conditions $\left| \boldsymbol{v}_{\mathrm{t}} \right| = 0$.
For arbitrary $\theta$, we obtain the general relation between the unknown and known quantities
\begin{equation}
\left| \boldsymbol{v}_{\perp 1} \right| + \left| \boldsymbol{v}_{\perp 2} \right| \frac{\overline{r}_2- \overline{r}_1 \cos \theta }{\overline{r}_1 - \overline{r}_2 \cos \theta} = \frac{\sin \theta \left( \left| \boldsymbol{v}_{\mathrm{l} 2} \right| \overline{r}_1 -  \left| \boldsymbol{v}_{\mathrm{l} 1} \right| \overline{r}_2 \right)}{\overline{r}_2 \cos \theta - \overline{r}_1} \;.
\label{eq:vanishing_v_t}
\end{equation}
Two special cases are
\begin{eqnarray}
\theta = 0 \; &\wedge& \; \left| \boldsymbol{v}_{\perp 1} \right| = \left| \boldsymbol{v}_{\perp 2} \right| \;, \label{eq:special1} \\
\left| \boldsymbol{v}_{\perp 1} \right| = \left| \boldsymbol{v}_{\perp 2} \right| \frac{\overline{r}_2 - \overline{r}_1 \cos \theta}{\overline{r}_2 \cos \theta - \overline{r}_1} \; &\wedge& \; \overline{r}_2 \left| \boldsymbol{v}_{\mathrm{l} 1} \right| = \overline{r}_1 \left| \boldsymbol{v}_{\mathrm{l} 2} \right|\;. \label{eq:special2}
\end{eqnarray}

Thus, the conditions for a purely radial infall are fine-tuned relative velocity components. 
The cases stated in prior work like Eq.~\eqref{eq:special1} within the measurement precision thus greatly depend on additional assumptions, like spherical symmetry, to achieve the fine-tuning. 
Only selecting galaxies almost aligned on the observer's line of sight is not sufficient to achieve $\left| \boldsymbol{v}_{\mathrm{t}} \right| =0$.

Assuming the perpendicular velocity components are unknown in Eq.~\eqref{eq:v_r_proj}, we find that the last term containing these components vanishes for $\theta \ne 0$ and $\left|\boldsymbol{r}_i \right| \ne 0$ if
\begin{equation}
\left| \boldsymbol{v}_{\perp 1} \right| \bar{r}_2 = \left| \boldsymbol{v}_{\perp 2} \right|\bar{r}_1 \; \Leftrightarrow \; \left( \frac{\left| \boldsymbol{v}_{\perp 1} \right|}{\left| \boldsymbol{v}_{\perp 2} \right|} = \frac{\left| \boldsymbol{r}_1 \right|}{\left| \boldsymbol{r}_2 \right|} \right) \; \vee \; \left( \boldsymbol{v}_{\perp 1} = \boldsymbol{v}_{\perp 2} = 0 \right)\;.
\label{eq:conditions}
\end{equation}

These insights have an important impact on the minor and major infall models as discussed below.

\subsection{Minor infall - projection on the connection line}
\label{sec:minor_infall}

\begin{figure}
\label{fig:minor_infall}
\centering
\includegraphics[width=0.4\textwidth]{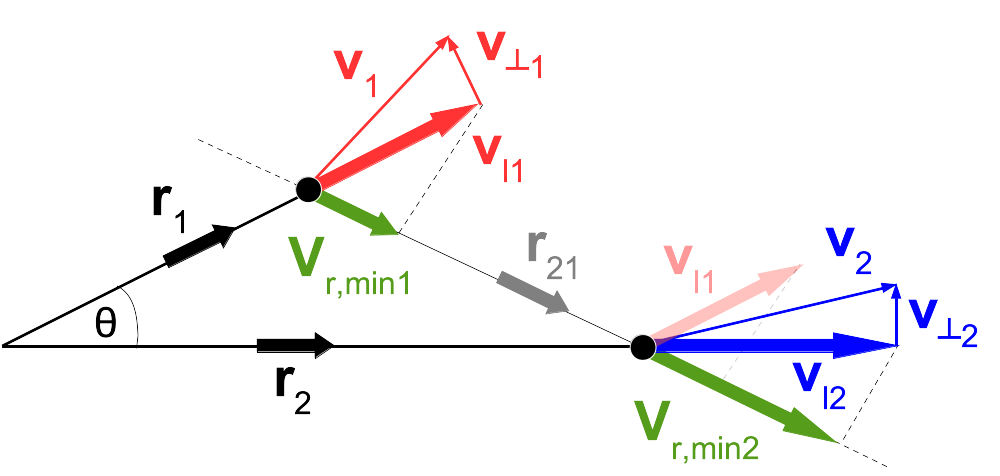}
\caption{Minor infall model: the line-of-sight velocity components of both galaxies are projected onto their connection line. The difference of these projections yields the relative radial velocity.}
\end{figure}

In this infall model, originally defined to describe the motion of two galaxies whose velocities are dominated by their surrounding Hubble flow, $\left| \boldsymbol{v}_{\mathrm{l} i} \right| \gg \left| \boldsymbol{v}_{\perp i} \right|$, their relative radial velocity is
\begin{align}
\left| \boldsymbol{v}_{\mathrm{r},\mathrm{min}} \right| &\equiv \left( \boldsymbol{v}_{\mathrm{l} 2} - \boldsymbol{v}_{\mathrm{l} 1} \right) \cdot \hat{\boldsymbol{r}}_{21} \\
&= \left| \boldsymbol{v}_{\mathrm{l} 1} \right| \overline{r}_1 + \left| \boldsymbol{v}_{\mathrm{l} 2} \right| \overline{r}_2 - \cos \theta  \left( \left| \boldsymbol{v}_{\mathrm{l} 1} \right| \overline{r}_2 + \left| \boldsymbol{v}_{\mathrm{l} 2} \right| \overline{r}_1 \right) \label{eq:v_min_proj} \;,
\end{align}
which neglects the last term in Eq.~\eqref{eq:v_r_proj}.
Fig.~\ref{fig:minor_infall} visualises the situation.
For $\theta \ne 0$ and $\boldsymbol{r}_i \ne 0$, we find that the minor infall model is a good approximation to the true radial velocity, if one of the two options in Eq.~\eqref{eq:conditions} holds.
We also read off Eqs.~\eqref{eq:v_r_proj} and \eqref{eq:v_min_proj} that, for $0 < \theta < \pi$, $\left| \boldsymbol{v}_{\mathrm{r}} \right|$ is under-estimated by Eq.~\eqref{eq:v_min_proj} if $\left| \boldsymbol{v}_{\perp 1} \right|/\left| \boldsymbol{v}_{\perp 2} \right| > \left| \boldsymbol{r}_1 \right|/\left| \boldsymbol{r}_2 \right|$.
Vice versa, $\left| \boldsymbol{v}_{\mathrm{r}} \right|$ is over-estimated by Eq.~\eqref{eq:v_min_proj} if $\left| \boldsymbol{v}_{\perp 1} \right|/\left| \boldsymbol{v}_{\perp 2} \right| < \left| \boldsymbol{r}_1 \right|/\left| \boldsymbol{r}_2 \right|$.

For $\boldsymbol{v}_i \approx \boldsymbol{v}_{\mathrm{l} i}, \, i=1,2$, meaning $\boldsymbol{v}_{\perp i} \approx 0$, we find that Eq.~\eqref{eq:v_t_proj} is reduced to the first term which consists of observable quantities up to the unknown direction of $\hat{\boldsymbol{r}}_{\perp 21}$.
So it is impossible to calculate $\boldsymbol{v}_{\mathrm{t}}$ in general, which explains why the minor infall model does not make a statement about it.
At best, we can exploit \fixme{the fact that we observe $\theta$ with} $\sin \theta \in \left[-1; 1 \right]$ for $\boldsymbol{v}_i \approx \boldsymbol{v}_{\mathrm{l} i}$ to constrain $\left| \boldsymbol{v}_{\mathrm{t}} \right|$ by
\begin{equation}
- \fixme{\lVert  \sin \theta \left( \left| \boldsymbol{v}_{\mathrm{l} 2} \right| \overline{r}_1 - \left| \boldsymbol{v}_{\mathrm{l} 1} \right| \overline{r}_2 \right) \rVert_2} \le \left| \boldsymbol{v}_{\mathrm{t}} \right| \le + \fixme{ \lVert \sin \theta \left( \left| \boldsymbol{v}_{\mathrm{l} 2} \right| \overline{r}_1 - \left| \boldsymbol{v}_{\mathrm{l} 1} \right| \overline{r}_2 \right) \rVert_2} \;.
\label{eq:vt_bound}
\end{equation}

\subsection{Major infall - projection on one line of sight}
\label{sec:major_infall}

We first employ this infall model for two galaxies.
Yet, we already note that the original approach replaced galaxy~2 by the centre of mass of a galaxy group and considered the infall of a galaxy onto the centre instead of a binary motion (see Sect.~\ref{sec:galaxy_clusters}). 

\begin{figure}
\label{fig:major_infall}
\centering
\includegraphics[width=0.35\textwidth]{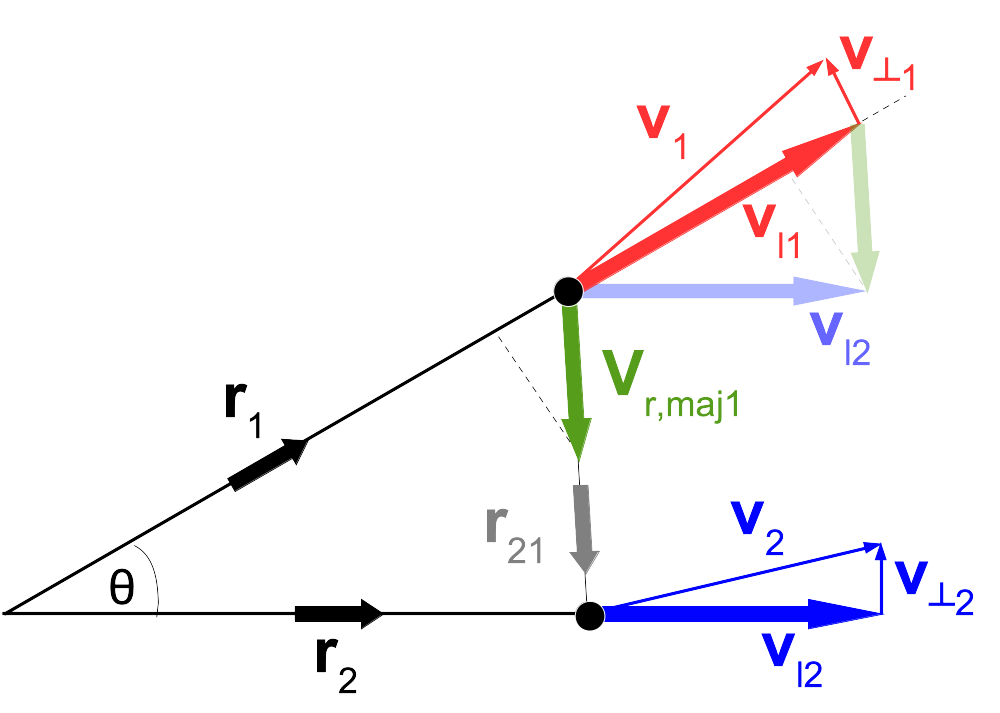}
\caption{Major infall model for galaxy~1: the radial infall velocity of galaxy~1 onto galaxy~2 is determined from the projection of the radial infall velocity and the line-of-sight velocity of galaxy~2 onto the line-of-sight of galaxy~1. An analogous procedure yields the major infall model for galaxy~2. Due to the asymmetry of this model, both radial infall velocities need not be of the same size.}
\end{figure}

Shown in Fig.~\ref{fig:major_infall}, the major infall model determines the radial velocity of galaxy~1 with respect to galaxy~2 by projecting all relevant velocity components onto $\hat{\boldsymbol{r}}_1$
\begin{equation}
\left| \boldsymbol{v}_{\mathrm{r}, \mathrm{maj} 1} \right| \hat{\boldsymbol{r}}_{21} \cdot \hat{\boldsymbol{r}}_1 = \left( \boldsymbol{v}_{\mathrm{l} 2} - \boldsymbol{v}_{\mathrm{l} 1} \right) \cdot \hat{\boldsymbol{r}}_1 \;,
\end{equation}
analogously for galaxy~2. 
Solving for $\left| \boldsymbol{v}_{\mathrm{r}, \mathrm{maj} i} \right|$ for $i=1,2$, yields
\begin{eqnarray}
\left| \boldsymbol{v}_{\mathrm{r}, \mathrm{maj} 1} \right| &=& \frac{\left| \boldsymbol{v}_{\mathrm{l} 2} \right| \cos \theta - \left| \boldsymbol{v}_{\mathrm{l} 1} \right|}{\overline{r}_2 \cos \theta -\overline{r}_1} \;, \label{eq:v_maj_1}\\
\left| \boldsymbol{v}_{\mathrm{r}, \mathrm{maj} 2} \right| &=& \frac{\left| \boldsymbol{v}_{\mathrm{l} 1} \right| \cos \theta - \left| \boldsymbol{v}_{\mathrm{l} 2} \right|}{\overline{r}_2 -\overline{r}_1 \cos \theta}\;. \label{eq:v_maj_2}
\end{eqnarray}
For comparison, the generalised version of the major infall model is obtained by projecting Eq.~\eqref{eq:v_rad_tan} onto $\hat{\boldsymbol{r}}_i$
\begin{equation}
\left( \boldsymbol{v}_2 - \boldsymbol{v}_1 \right) \cdot \hat{\boldsymbol{r}}_i = \left( \left| \boldsymbol{v}_{\mathrm{r}}\right| \hat{\boldsymbol{r}}_{21} + \left| \boldsymbol{v}_{\mathrm{t}} \right| \hat{\boldsymbol{r}}_{\perp 21} \right) \cdot \hat{\boldsymbol{r}}_i \;, \quad i=1,2 \;.
\end{equation}
Solving for $\left| \boldsymbol{v}_{r} \right|$, we obtain
\begin{eqnarray}
\left| \boldsymbol{v}_{\mathrm{r}} \right|  &=& \frac{\left| \boldsymbol{v}_{\mathrm{l} 2} \right| \cos \theta - \left| \boldsymbol{v}_{\mathrm{l} 1} \right| + \sin \theta \left( \left| \boldsymbol{v}_{\perp 2} \right| - \overline{r}_2 \left| \boldsymbol{v}_{\mathrm{t}} \right| \right)}{\left(\overline{r}_2 \cos \theta - \overline{r}_1 \right)} \;, \; i=1 \label{eq:v_maj_1_full} \\
&\approx& \left| \boldsymbol{v}_{\mathrm{l} 2} \right| - \left| \boldsymbol{v}_{\mathrm{l} 1} \right| + \theta \left( \left| \boldsymbol{v}_{\perp 2} \right| - \overline{r}_2 \left| \boldsymbol{v}_{\mathrm{t}} \right| \right) \;, \\
\left| \boldsymbol{v}_{\mathrm{r}} \right| &=& \frac{\left| \boldsymbol{v}_{\mathrm{l} 2} \right| - \left| \boldsymbol{v}_{\mathrm{l} 1} \right| \cos \theta + \sin \theta \left( \left| \boldsymbol{v}_{\perp 1} \right| - \overline{r}_1 \left| \boldsymbol{v}_{\mathrm{t}} \right| \right)}{\left(\overline{r}_2 - \overline{r}_1 \cos \theta \right)} \;, \; i=2 \label{eq:v_maj_2_full} \\
&\approx& \left| \boldsymbol{v}_{\mathrm{l} 2} \right| - \left| \boldsymbol{v}_{\mathrm{l} 1} \right| + \theta \left( \left| \boldsymbol{v}_{\perp 1} \right| - \overline{r}_1 \left| \boldsymbol{v}_{\mathrm{t}} \right| \right) \;.
\end{eqnarray}
Thus, the major infall model is accurate for $\theta \ne 0$ and $\boldsymbol{r}_i \ne 0$ if
\begin{eqnarray}
i=1: \; \left| \boldsymbol{v}_{\perp 2} \right| = \overline{r}_2 \left| \boldsymbol{v}_{\mathrm{t}} \right| \;, \quad \text{or} \quad 
i=2: \; \left| \boldsymbol{v}_{\perp 1} \right| = \overline{r}_1 \left| \boldsymbol{v}_{\mathrm{t}} \right|  \;.
\label{eq:v_maj_cond} 
\end{eqnarray}
Combining both equations to eliminate the unknown $\left| \boldsymbol{v}_{\mathrm{t}} \right|$ yields Eq.~\eqref{eq:conditions} again.
So, the conditions for over- or under-estimation of the true radial velocity are the same as for the minor infall.

A special case is $\left| \boldsymbol{v}_{\mathrm{t}} \right| =0$, implying a purely radial infall, and $\left| \boldsymbol{v}_{\perp i} \right| =0$, $i=1,2$. 
Only considering one equation of Eqs.~\eqref{eq:v_maj_1_full} and \eqref{eq:v_maj_2_full}, this case yields a purely radial infall of one galaxy under a symmetry constraint that forces the perpendicular velocity of the other galaxy or the group centre to vanish (see Sect.~\ref{sec:galaxy_clusters}).  

\subsection{Comparison and limits}
\label{sec:comparison}

The general versions of the infall models are easily related to each other via the different projections and it is obvious that both models are independent of a background cosmology. 
The minor infall model is best used when we do not have more information beyond $\theta$, $\boldsymbol{v}_{\mathrm{l} i}$, $i=1,2$. 
Since the model does not make any assumption about $\boldsymbol{v}_{\mathrm{t}}$ and treats each galaxy of a pair equally, it is well-suitable to describe binary galaxies. 

In contrast, the major infall model is best used for an ensemble of several galaxies. 
While it can also describe a pair, it requires information about the tangential and perpendicular components.
The latter can be added via symmetry assumptions, readily testable in groups but hard to establish for binaries, \fixme{with the Milky Way and Andromeda system being the only exception known so far, \cite{bib:Benisty2022}}. 

Replacing galaxy~2 by the centre of mass of the entire structure, explains why the major infall model is asymmetric in the infalling objects. 
Any external impact, like an accelerated cosmic expansion, will not generate additional angular momentum, increasing $\boldsymbol{v}_{\perp 2}$.
Symmetry assumptions and a correspondingly well-sampled distribution of galaxies then allow us to describe groups and clusters robustly even over cosmic time.

Comparing Eqs.~\eqref{eq:v_min_proj} and \eqref{eq:v_maj_1}, or \eqref{eq:v_maj_2},
their difference reads
\begin{eqnarray}
\left| \boldsymbol{v}_{\mathrm{r}, \mathrm{min}} \right| -\left| \boldsymbol{v}_{\mathrm{r}, \mathrm{maj} 1} \right| = \fixme{ \tfrac{\overline{r}_2 \left( \cos^2 \theta -1 \right)}{\overline{r}_2 \cos \theta - \overline{r}_1} \left( \left| \boldsymbol{v}_{\mathrm{l} 2} \right| \overline{r}_1 - \left| \boldsymbol{v}_{\mathrm{l} 1} \right| \overline{r}_2 \right)}
\label{eq:equality}
\end{eqnarray}
and analogously for $\left| \boldsymbol{v}_{\mathrm{r}, \mathrm{maj} 2} \right|$. 
For $\boldsymbol{r}_i \ne 0$ and \fixme{$\boldsymbol{v}_{\mathrm{l} 2} \overline{r}_1 \ne \boldsymbol{v}_{\mathrm{l} 1} \overline{r}_2$}, the infall models only agree for $\theta = 0$ \fixme{or $\theta = \pi$}. Thus, for small angles, as considered in prior work, both models coincide in the trivial relation up to small deviations 
\begin{equation}
\left| \boldsymbol{v}_{\mathrm{r}, \mathrm{min}} \right| = \left| \boldsymbol{v}_{\mathrm{r}, \mathrm{maj} 1} \right| = \left| \boldsymbol{v}_{\mathrm{r}, \mathrm{maj} 2} \right| = \left| \boldsymbol{v}_\mathrm{r} \right| = \left| \boldsymbol{v}_{\mathrm{l} 2} \right| - \left| \boldsymbol{v}_{\mathrm{l}1} \right| \;.
\label{eq:theta0}
\end{equation}
\fixme{In addition, for $\theta = \pi$, both infall models coincide again.}
For $0 < \theta < \pi$, \fixme{the difference can be positive or negative depending on the relations between all quantities involved.} 
Since both models \fixme{can} over- or under-estimate the true radial infall velocity based on Eq.~\eqref{eq:conditions}, we cannot decide which model is closer to the true radial component without additional information or assumptions.

\section{From binaries to groups and clusters}
\label{sec:galaxy_clusters}

The approaches in \cite{bib:Karachentsev2006} rely on spherical symmetry and require the member galaxies to sample the total spherical mass distribution well. 
To investigate the impact of each assumption, let the centre of the \emph{total} mass distribution (stellar, gaseous, dark mass) be $\boldsymbol{r}_{\mathrm{cm}}$ and its total velocity $\boldsymbol{v}_{\mathrm{cm}} = \rm{d}\boldsymbol{r}_{\mathrm{cm}}/\rm{d}t$.
For $n$ member galaxies sampling this mass-density profile, their centre of mass and its velocity are given by
\begin{equation} 
\boldsymbol{r}_{\mathrm{cg}} = \frac{1}{M_{\mathrm{g}}} \sum \limits_{i=1}^{n}m_i \boldsymbol{r}_i \;, \quad
\boldsymbol{v}_{\mathrm{cg}} = \frac{1}{M_{\mathrm{g}}} \sum \limits_{i=1}^{n}m_i \boldsymbol{v}_i \;,
\label{eq:cm}
\end{equation}
where $M_{\mathrm{g}}$ is the mass of all galaxies and we assumed that the galaxies move non-relativistically.
While $M_\mathrm{g}$ is known when the mass-to-light ratio of the galaxies is known, the total mass of the structure $M$ remains unknown, for instance, including a galaxy-cluster-scale dark-matter halo in which the galaxies are moving, so $M \ge M_{\mathrm{g}}$. 
The galaxies are a representative sample if, at least, $\boldsymbol{r}_{\mathrm{cm}}=\boldsymbol{r}_{\mathrm{cg}}$ and $\boldsymbol{v}_{\mathrm{cm}}=\boldsymbol{v}_{\mathrm{cg}}$. 
These quantities need not agree generally, for instance, for a very small number of galaxies, in merger scenarios, or if the dark matter distribution is not well-traced by the luminous matter.
For $\boldsymbol{r}_i = \boldsymbol{r}_{\mathrm{cm}} - \boldsymbol{r}_{\mathrm{cm} \, i}$ and $\boldsymbol{v}_i = \boldsymbol{v}_{\mathrm{cm}} - \boldsymbol{v}_{\mathrm{cm} \, i}$,
the total momentum with respect to the observer is
\begin{equation}
\boldsymbol{P} = M_\mathrm{g} \boldsymbol{v}_\mathrm{cm} - \sum \limits_{i=1}^{n} m_i \, \boldsymbol{v}_{\mathrm{cm} \, i} \;.
\label{eq:P}
\end{equation}
If $\boldsymbol{r}_{\mathrm{cm}}=\boldsymbol{r}_{\mathrm{cg}}$, the sum vanishes.
Assuming a spherically symmetric structure on a homogeneous and isotropic background, $\boldsymbol{v}_\mathrm{cm}$ reduces to the cosmic velocity along the observer's line of sight.
Without these assumptions, $\boldsymbol{P}$ is not well-constrained even for $n\gg2$ due to the unknown perpendicular and tangential parts.

The total angular momentum with respect to the observer is
\begin{eqnarray}
\boldsymbol{L} = \sum \limits_{i=1}^{n} m_i \, \boldsymbol{r}_i \times \boldsymbol{v}_i \equiv \left| \boldsymbol{L}_1 + \boldsymbol{L}_2 - \boldsymbol{L}_3 - \boldsymbol{L}_4 \right| \, \hat{\boldsymbol{r}}_{\mathrm{cm}} \times \hat{\boldsymbol{r}}_{\perp \mathrm{cm}} \label{eq:L4} \;. 
\end{eqnarray}
With the relations in Sect.~\ref{sec:definitions}, the four parts read
\begin{eqnarray}
\left| \boldsymbol{L}_1 \right| &=& \Big| \sum \limits_{i=1}^{n} m_i \, \boldsymbol{r}_{\mathrm{cm}} \times \boldsymbol{v}_{\perp \mathrm{cm}} \Big| = M_{\mathrm{g}} \left| \boldsymbol{r}_{\mathrm{cm}} \right| \left| \boldsymbol{v}_{\perp \mathrm{cm}} \right| \;, \label{eq:L_first}\\ 
\left| \boldsymbol{L}_2 \right| &=& \Big| \sum \limits_{i=1}^{n} m_i \, \boldsymbol{r}_{\mathrm{cm} \, i} \times \boldsymbol{v}_{\perp \mathrm{cm}\, i} \Big| = \sum \limits_{i=1}^{n} m_i \left|\boldsymbol{r}_{\mathrm{cm} \, i} \right| \left| \boldsymbol{v}_{\mathrm{t}\, i} \right| \;, 
\end{eqnarray}
\begin{eqnarray}
\left| \boldsymbol{L}_3 \right| &=& \Big| \sum \limits_{i=1}^{n} m_i \, \boldsymbol{r}_{\mathrm{cm} \, i} \times \boldsymbol{v}_{\mathrm{cm}} \Big| \;, \\
&=& \sum \limits_{i=1}^{n} m_i \, \Big( \left| \boldsymbol{v}_{\perp \mathrm{cm}} \right| (\left| \boldsymbol{r}_{\mathrm{cm}} \right| - \left|\boldsymbol{r}_i \right| \cos \theta_i) + \left| \boldsymbol{v}_{\mathrm{l} \, \mathrm{cm}} \right| \left| \boldsymbol{r}_i \right| \sin \theta_i \Big) \,, \\
\left| \boldsymbol{L}_4 \right| &=& \Big| \sum \limits_{i=1}^{n} m_i \, \boldsymbol{r}_{\mathrm{cm}} \times \boldsymbol{v}_{\mathrm{cm} \, i} \Big| = - \sum \limits_{i=1}^{n} m_i \, \left|\boldsymbol{v}_{\mathrm{r} \, i} \right| \left| \boldsymbol{r}_{\mathrm{cm}} \right| \overline{r}_i \sin \theta_i \;.
\label{eq:L_fourth}
\end{eqnarray}
For an arbitrary observer in a homogeneous and isotropic universe, there is no reason for a total angular momentum to exist, such that $\left| \boldsymbol{L} \right| = 0$ for an isolated single structure, supported by observations \cite{bib:Hawking1969,bib:Saadeh2016}. 
As assumed in \cite{bib:Karachentsev2006}, if $\boldsymbol{r}_{\mathrm{cm}} = \boldsymbol{r}_{\mathrm{cg}}$, $\boldsymbol{L}_3$ and $\boldsymbol{L}_4$ vanish.
From the remaining $\left| \boldsymbol{L}_1 \right| = -\left| \boldsymbol{L}_2 \right|$, we obtain  
\begin{equation}
\left| \boldsymbol{v}_{\perp \mathrm{cm}} \right| + \frac{m_j}{M_\mathrm{g}} \frac{\left| \boldsymbol{r}_{\mathrm{cm} \, j} \right|}{\left| \boldsymbol{r}_{\mathrm{cm}}\right|} \left| \boldsymbol{v}_{\mathrm{t} \, j} \right| = - \sum \limits_{i=1, i\ne j}^{n} \frac{m_i}{M_\mathrm{g}} \frac{\left| \boldsymbol{r}_{\mathrm{cm} \, i} \right|}{\left| \boldsymbol{r}_{\mathrm{cm}} \right|} \left| \boldsymbol{v}_{\mathrm{t} \, i} \right| \;,
\label{eq:L_maj}
\end{equation} 
taking galaxy~$j$ outside the sum, assuming it falls onto a structure of $n-1$ galaxies. 
Then, we require $n \gg 1$ and a spherically symmetric structure in which the right-hand side of Eq.~\eqref{eq:L_maj} averages to zero.
Eq.~\eqref{eq:L_maj} amounts to $\left| \boldsymbol{v}_{\perp \mathrm{cm}} \right| = \lambda \overline{r}_{\mathrm{cm}} \left| \boldsymbol{v}_{\mathrm{t} \, j} \right|$ compared to Eq.~\eqref{eq:v_maj_cond}.
However, only the projection of $\boldsymbol{v}_\mathrm{t}$ on the observer's sky can be observed.
So we may absorb $\lambda$ into the definition of $\boldsymbol{v}_\mathrm{t}$ for the infall model and study its degeneracies with the projection angle on the observer's sky in a subsequent step. 
The major infall model then relates $\left| \boldsymbol{v}_{\mathrm{t} \, j} \right|$ to $\left| \boldsymbol{v}_{\perp \mathrm{cm}} \right|$.

Alternatively, all galaxies are in the spherically symmetric structure on the right-hand side of Eq.~\eqref{eq:L_maj}, hence $\left| \boldsymbol{v}_{\perp \mathrm{cm}} \right|=0$. 
Inserting the latter into Eqs.~\eqref{eq:v_maj_1_full} and \eqref{eq:v_maj_2_full}, solving for $\left| \boldsymbol{v}_{\perp i} \right|$ and $\left| \boldsymbol{v}_{\mathrm{t} \, i} \right|$ yields
\begin{eqnarray}
\left| \boldsymbol{v}_{\perp i} \right| = 0 \quad \vee \quad \theta = 0 \quad \vee \quad \theta = \pi/2 \quad \vee \quad \theta = \pi \;. 
\end{eqnarray}
Hence, the $\left| \boldsymbol{v}_{\mathrm{t} \, i} \right|$ do not need to vanish, so that the minor infall model is equally suitable to be used in this case.

The trivial case $\left| \boldsymbol{v}_{\perp \mathrm{cm}} \right| = \left| \boldsymbol{v}_{\mathrm{t} \, j} \right| = 0$ mentioned in Sect.~\ref{sec:major_infall} is obtained for a purely radial infall, $\left| \boldsymbol{v}_{\mathrm{t} \, j} \right| = 0$ either for galaxy~j onto a spherical structure, or even $\forall j=1,...n$, both yielding $\left| \boldsymbol{L}_1\right| = \left| \boldsymbol{L}_2 \right| = 0$.
Deviations from Eq.~\eqref{eq:v_maj_cond} can already arise from $\left| \boldsymbol{L}_2 \right|$ due to a less symmetric structure or insufficient sampling $n$, apart from asymmetries between $\boldsymbol{r}_{\mathrm{cm}}$ and $\boldsymbol{r}_{\mathrm{cg}}$ causing $\boldsymbol{L}_3$ and $\boldsymbol{L}_4$ to be non-zero. 

Rewriting Eq.~\eqref{eq:L4} in terms of all $\left| \boldsymbol{v}_{\mathrm{r,maj} \, i} \right|$, we use Eq.~\eqref{eq:v_maj_1_full} with a $\lambda$-factor in front of $\left| \boldsymbol{v}_{\mathrm{t} \, i}\right|$ as above but without the mass ratio, and $\theta_i \ne 0$ to obtain 
\begin{equation}
\left| \boldsymbol{L}_1 + \boldsymbol{L}_2 \right| = \sum \limits_{i=1}^{n} \frac{m_i}{\sin \theta_i} \left| \boldsymbol{r}_{\mathrm{cm}} \right| (\overline{r}_{\mathrm{cm}} \cos \theta_i - \overline{r}_i) \left( \left| \boldsymbol{v}_{\mathrm{r} \, i} \right| - \left| \boldsymbol{v}_{\mathrm{r,maj} \, i} \right| \right) \;.
\label{eq:L1L2}
\end{equation}
For $\theta_i = 0$, Eq.~\eqref{eq:theta0} holds without giving a contribution to Eq.~\eqref{eq:L1L2}. 
Detailed in Sect.~\ref{sec:major_infall}, the major infall model may over- or under-estimate $\left| \boldsymbol{v}_\mathrm{r} \right|$, such that any $\left| \boldsymbol{L} \right| \ne 0$ depends on the sum of deviations for all galaxies, which, in turn, depends on the symmetry of the structure. 
Since the minor infall does not include $\left| \boldsymbol{v}_\mathrm{t} \right|$, Eq.~\eqref{eq:L1L2} only relates the major infall model to the true radial velocity.

\section{\fixme{Application to simulated structures}}
\label{sec:simulation}

\fixme{
As proof-of-principle test, we apply the infall models to the snapshot of cosmic structures at $z=0$ in the Illustris-3 simulation, \cite{bib:Vogelsberger2014} and \cite{bib:Nelson2015}, which is based on our cosmological concordance model.
While Sect.~\ref{sec:pairwise_infall} tackles binaries and groups with more than two members alike, \cite{bib:Benisty2025a} will provide more details about infall models for binaries, so that we focus on groups here. 
}

\fixme{
From all halos identified by the friends-of-friends clustering, we select those with a galaxy-group-scale mass $m_\mathrm{halo} > 10^{12}~M_\odot/h$ with the dimensionless Hubble constant $h=0.704$.
To ensure that all halos and their subhalos are well-sampled, we consider halos with a minimum number of particles larger than 100 and subhalos with at least 20 particles. 
While Illustris-3 is the least resolved simulation compared to Illustris-1 and~2, its halo and subhalo catalogues list the centre of mass position of the halos and subhalos in addition to the position of the most bound particle. 
The additional information allows us to select relaxed halos and non-merging subhalos: we require both comoving positions to differ by less than 10~ckpc$/h$ for halos and subhalos. 
Yet, this is just one option among many, as detailed in \cite{bib:Zjupa2017} and references therein. 
}

\fixme{
We furthermore require that the halos are isolated, meaning we only investigate halos without any neighbouring halos within the radius of their own zero-velocity surface $r_0 \equiv \lVert\boldsymbol{r}_0\rVert_2$ plus the maximum $r_0$ of all halos in the simulation. 
To calculate $r_0$, we use the relation derived in \cite{bib:Peirani2006}
\begin{equation}
m_{0} \approx 4.1 \times 10^{12} h^2 \left(\frac{r_0}{1~\mbox{Mpc}}\right)^3~M_\odot
\label{eqn:m0}
\end{equation}
for $r_0$, setting the mass enclosed in this radius $m_0$ equal to the simulated halo mass.
For our final set of 344 halos, $r_0 \in \left[0.89, 2.00 \right]~\mbox{Mpc}$. 
As we assume that the origin of the simulation is a random position in the snapshot, we place the observer at this position. 
Fig.~\ref{fig:halo_dist} shows the distribution of physical distances versus the halo masses for the 344 selected halos. 
\begin{figure}
\centering
\includegraphics[width=0.9\linewidth]{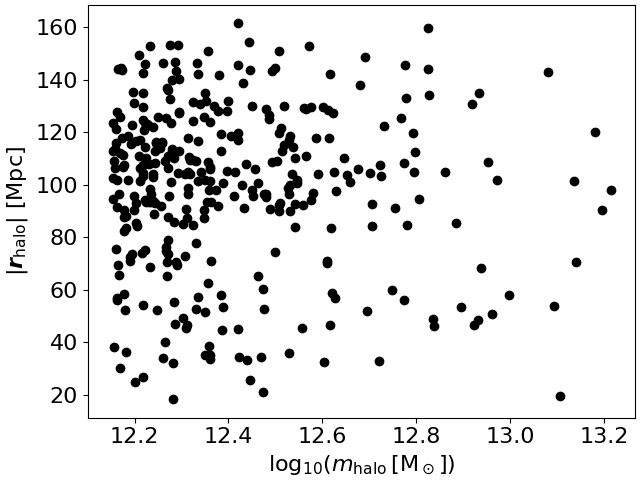}
\caption{\fixme{Physical distances from an observer at the origin to the centre of mass of 344 isolated, relaxed halos versus their halo masses in the $z=0$-snapshot of Illustris-3. In the notation of Sect.~\ref{sec:galaxy_clusters}, $|\boldsymbol{r}_\mathrm{halo}| \equiv |\boldsymbol{r}_\mathrm{cm}|$.}}
\label{fig:halo_dist}
\end{figure}
}

\fixme{
For the selection of subhalos, we pursue two approaches: first, we load all subhalos that are gravitationally bound to their parent halo as indicated in the subhalo catalogue. 
Second, to study the behaviour of the infall models in the Hubble flow around the structures, we select all subhalos that are within a sphere of $1.5\, r_0$. 
For the first approach, the number of subhalos per halo is $n_\mathrm{sub} \in \left[1,38 \right]$ with a total of 1476 bound subhalos, for the second $n_\mathrm{sub} \in \left[1, 92 \right]$ with a total of 5036 subhalos.
}

\fixme{
\begin{figure*}
\centering
\includegraphics[width=0.33\linewidth]{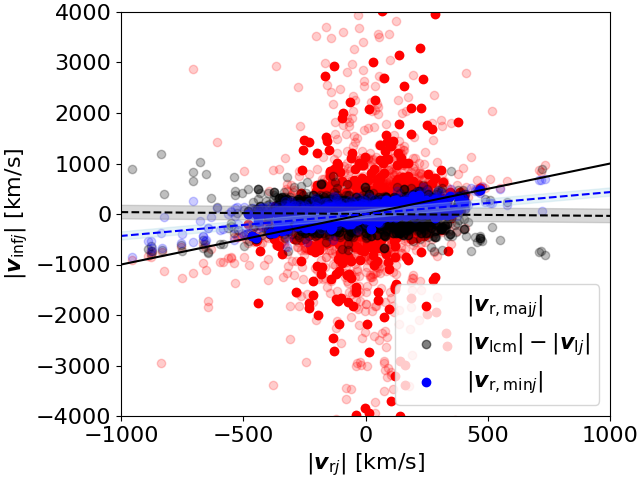} \hfill
\includegraphics[width=0.33\linewidth]{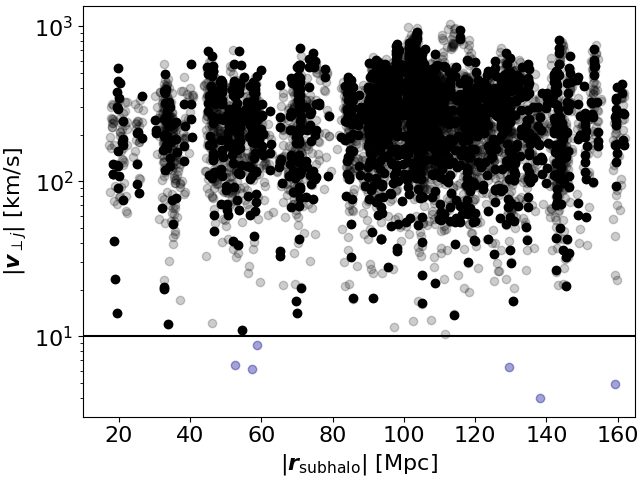} \hfill
\includegraphics[width=0.33\linewidth]{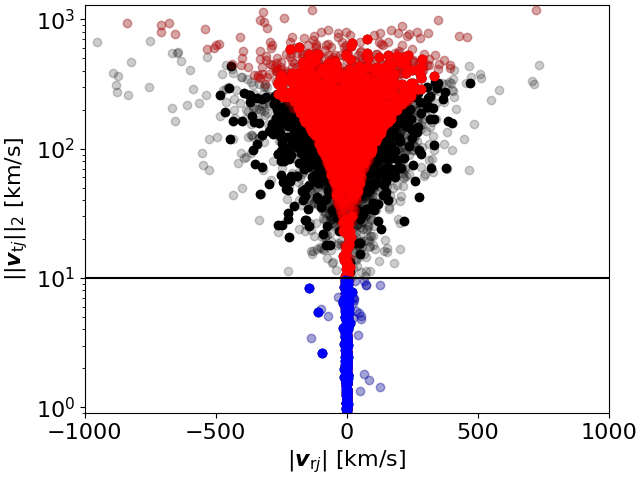}
\caption{\fixme{Left: Reconstructed infall velocities by Eqs.~\eqref{eq:v_min_proj}, \eqref{eq:v_maj_1}, and \eqref{eq:theta0} for all subhalos (bound in dark colours, Hubble-flow subhalos in light colours) onto their parent-halo centre versus the true radial velocity by Eq.~\eqref{eq:v_r_proj}. Linear fits with 1-$\sigma$ confidence bounds to the point clouds of the minor infall model and the velocity-difference approximation are shown in blue and grey, respectively. The fit for the major infall is not shown as its confidence bound covers the entire plot. Centre: Perpendicular velocities of all subhalos versus their distance to the observer (bound in black, Hubble-flow subhalos in grey). Only the six subhalos in the Hubble flow marked in blue have a tangential velocity close to zero. In the notation of Sect.~\ref{sec:galaxy_clusters}, $|\boldsymbol{r}_\mathrm{subhalo}| \equiv |\boldsymbol{r}_j|$. Right: Absolute value of the tangential velocity of all subhalos (bound in dark colours, Hubble-flow subhalos in light colours) versus their radial velocity. Only 337 subhalos with radial velocity close to zero also have a tangential velocity close to zero (marked in blue), while 3575 subhalos have a larger tangential velocity than radial one (marked in red).}}
\label{fig:vinf_vrad}
\end{figure*}
After the selection process, we convert all comoving quantities to physical ones.
This requires to add a Hubble flow to the simulated velocities of all halos and subhalos $\boldsymbol{v}_{\mathrm{sim} j}$ as
\begin{equation}
\boldsymbol{v}_j = \boldsymbol{v}_{\mathrm{sim} j} + H_0 \boldsymbol{r}_j .    
\end{equation}
As a result, we obtain the final data on which we perform our infall-model tests.
}

\fixme{
While the position of the most bound particle is usually used to locate halos and subhalos, we employ the centre of mass coordinates instead. 
This is motivated by the derivations of Sect.~\ref{sec:galaxy_clusters}.
Based on our selection criteria, the differences between using the most bound particle or the centre of mass are minor. 
Moreover, using the most bound particle, numerical instabilities occur for most of the central subhalos when their most bound particle has \emph{exactly} the same position as the most bound particle of the parent halo, such that no infall models can be calculated for it. 
}

\fixme{
At first, we test the accuracy of the minor and major infall models to reconstruct the radial velocity. 
Since the halos are at distances beyond 18~Mpc, $\lVert \boldsymbol{r}_{\mathrm{cm} j}\rVert_2 \in \left[0, 3 \right]~\mbox{Mpc}$ with $\lVert \theta \rVert_2 \in \left[0, 7.4 \right]~\mbox{deg}$ and thus small compared to the angles of objects in our cosmic neighbourhood investigated by \cite{bib:Karachentsev2006}, \cite{bib:Karachentsev2010}, and others. 
Thus, the small-angle approximation discussed in Sect.~\ref{sec:pairwise_infall} should hold, implying that the $\sin \theta$-terms in Eqs.~\eqref{eq:v_r_proj}, \eqref{eq:v_maj_1_full}, and \eqref{eq:v_maj_2_full} are suppressed. 
}

\fixme{
Yet, Fig.~\ref{fig:vinf_vrad} shows a different picture: In Fig.~\ref{fig:vinf_vrad} (left), we plot the infall velocities according to Eqs.~\eqref{eq:v_min_proj}, \eqref{eq:v_maj_1}, and \eqref{eq:theta0} of all 5036 subhalos, bound and unbound, onto their parent-halo centre versus the true radial velocity. 
The approximation of Eq.~\eqref{eq:theta0} shows a similar scatter as the minor infall model, supporting the statement in Sect.~\ref{sec:comparison} that it is a good approximation of the latter for the small angles at least out to 7.4~deg. 
Both models have a limited spread, such that a linear fit $|\boldsymbol{v}_{\mathrm{inf} j}| = m_\mathrm{v} |\boldsymbol{v}_{\mathrm{r} j}|$ can be calculated together with its 1-$\sigma$ confidence bounds $b_\mathrm{v}$. 
For the minor infall model we obtain $m_\mathrm{v}=0.43$, $b_\mathrm{v}=81.06$~km/s, for the approximation by Eq.~\eqref{eq:theta0} $m_\mathrm{v}=-0.04$, $b_\mathrm{v}=153.39$~km/s.
Comparing Eqs.~\eqref{eq:v_min_proj} and \eqref{eq:v_maj_1}, the spread of the major infall velocity is larger, which is caused by the fact that the major infall model is a ratio and the denominator can have a small absolute value.
Fitting a line to this data, yields $m_v=1.01$, but also $b_\mathrm{v}=5578.32$~km/s.
As Fig.~\ref{fig:vinf_vrad} (left) only shows infall velocities up to an absolute value of 4000~km/s, not all major-infall velocities are in the plot.
Restricting the line fits to consider bound subhalos only, the values only change marginally.
}

\fixme{
To investigate the source of the deviations from the true radial velocity, we plot the perpendicular velocities for each subhalo versus its physical distance to us in Fig.~\ref{fig:vinf_vrad} (centre). 
As the plot shows, there is no trend of vanishing perpendicular velocities visible, although they maximally amount to 34\% of the line-of-sight velocities and are therefore much smaller. 
Only six subhalos in the Hubble flow have perpendicular velocities smaller than 10~km/s, which can be interpreted as conservatively compatible with zero, see for instance \cite{bib:Kim2020}.  
So even though $\sin \theta$ is small, the perpendicular velocities multiplied by the normalised distances compensate for the small angle and therefore yield a non-negligible contribution to the last term of the radial infall velocity in Eq.~\eqref{eq:v_r_proj}. 
This is in contrast to the assumptions made in \cite{bib:Karachentsev2006} that the peculiar velocities of the galaxies are small compared to the line-of-sight velocities subject to the Hubble flow. 
}

\fixme{
Since the perpendicular velocities are not negligible, we plot the absolute value of the tangential velocity for all subhalos in Fig.~\ref{fig:vinf_vrad} (right). 
The plot shows a similar result. 
Only 337 subhalos out of 5036 subhalos have a tangential velocity that is smaller than 10~km/s. 
Further investigating the properties of these subhalos, we discover that 309 approximately obey Eq.~\eqref{eq:special1} with $\theta \approx 0$, even $|\boldsymbol{r}_{\mathrm{cm}j}| \approx 0$ and a maximum $||\boldsymbol{v}_{\perp \mathrm{cm}} - \boldsymbol{v}_{\perp j}||_2 = 12$~km/s. 
These subhalos are the central subhalos close to the centre of the parent halo. 
For 3 additional bound halos and 25 in the Hubble flow, another fine-tuning reduces $||\boldsymbol{v}_{\mathrm{t}j}||_2<10$~km/s by chance.
The remaining 4699 subhalos thus have a non-negligible tangential velocity, for 3575 of those even $||\boldsymbol{v}_{\mathrm{t}j}||_2 > ||\boldsymbol{v}_{\mathrm{r} j}||_2$ holds.
In summary, 93\% of all tangential velocities are not negligible and 71\% of all $||\boldsymbol{v}_{\mathrm{t}j}||_2$ exceed $||\boldsymbol{v}_{\mathrm{r}j}||_2$.
}

\fixme{
Based on these findings and supported by Eq.~\eqref{eq:special1}, it is not sufficient to restrict the application of the infall models to galaxies that are in front and behind the cluster centre with $\theta \approx 0$ in order to reduce the impact of the tangential velocity, as stated in \cite{bib:Karachentsev2010}. 
Moreover, \cite{bib:Karachentsev2006} studied the impact of $\boldsymbol{v}_{\mathrm{t}j}$ by simulating a distributions of tangential velocities as a Gaussian amplitude with mean 70~km/s and standard deviation of 30~km/s combined with a uniform distribution in the orientation. 
While this might be a realistic choice of parameters for the close-by low-mass cosmic structures they investigate, the result does not hold in general, as Fig.~\ref{fig:vinf_vrad} (right) reveals.}

\fixme{ 
\begin{figure*}
\centering
\includegraphics[width=0.33\linewidth]{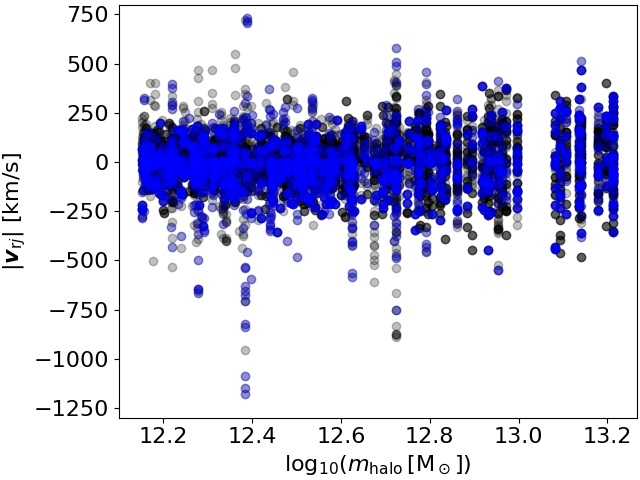} \hfill
\includegraphics[width=0.33\linewidth]{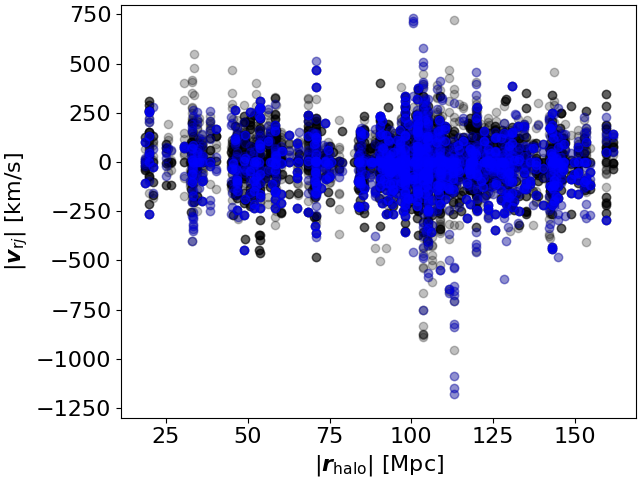} \hfill
\includegraphics[width=0.33\linewidth]{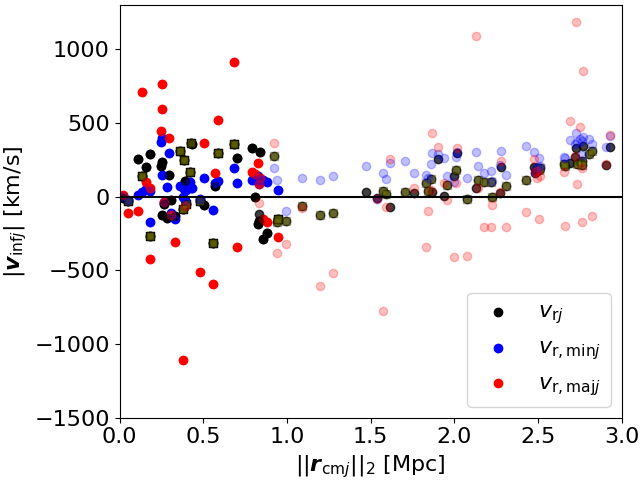}
\caption{\fixme{Left: Radial velocity of all 5036 subhalos onto their parent halo versus the mass of their parent halo, highlighted in blue are those subhalos whose radial velocity lies between the minor and major infall velocity (dark colours mark bound halos, light colours mark subhalos in the Hubble flow). Centre: Same as the previous plot, but depending on the distance of the parent halo to the observer instead of the parent-halo mass. Right: Hubble-flow diagram for the most massive halo in our halo set to be compared with Fig.~7 of \cite{bib:Kim2020}. The major and minor infall velocities can be compared to the true radial infall velocity (dark colours mark bound halos, light colours mark subhalos in the Hubble flow), squared markers indicate which radial velocities lie between the minor and major infall velocities.}}
\label{fig:Kim_discussion}
\end{figure*}
At last, we investigate the statement made in \cite{bib:Kim2020} that the true radial velocities of the Virgo cluster are expected to lie between the major and minor infall model estimates, which is not supported by the theoretical derivations in Sect.~\ref{sec:comparison}. 
Out of 5036 subhalos in our data set, only 34\%, meaning 1707, fulfill this condition. 
Since Virgo is at 16.5~Mpc from us as observers and has $m_0 \approx 6 \times 10^{14}~M_\odot$ according to \cite{bib:Kim2020}, it is out of the distance and mass range covered by our halo selection.
However, the infall models are based on a purely geometrical kinematics construct, so that it is very likely that our findings also apply to the Virgo cluster. 
To support this claim, Fig.~\ref{fig:Kim_discussion} (left, centre) shows there is no correlation between the radial velocity of a subhalo and its parent halo mass or the physical distance of its parent halo to the observer. 
Marking all subhalos whose radial velocity lies between the minor and major infall velocity in blue, we also read off Fig.~\ref{fig:Kim_discussion} (left, centre) that there is no correlation for those cases, either.
Analogously to Fig.~7 in \cite{bib:Kim2020}, we plot the Hubble flow around the centre of our most massive halo in Fig.~\ref{fig:Kim_discussion} (right) to show that the true radial velocity need not lie between the minor and major infall velocity. 
For this halo, $m_\mathrm{halo}=1.64 \times 10^{13}~M_\odot$ corresponding to $r_0 = 2.00$~Mpc. 
We count 38 bound subhalos and 54 in the Hubble flow out to $1.5\, r_0$, amounting to 92 subhalos in total.
For those, 14 out of 38 bound subhalos have a radial velocity between the minor and major infall velocities, and 31 out of the 54 in the Hubble flow. 
Hence, only 49\% of all subhalos fulfil this constraint. 
(For the second most massive halo with a total of 43 subhalos (20 bound, 23 in the Hubble flow), the ratio is only 30\%.)
}

\fixme{
\begin{figure}
\centering
\includegraphics[width=0.9\linewidth]{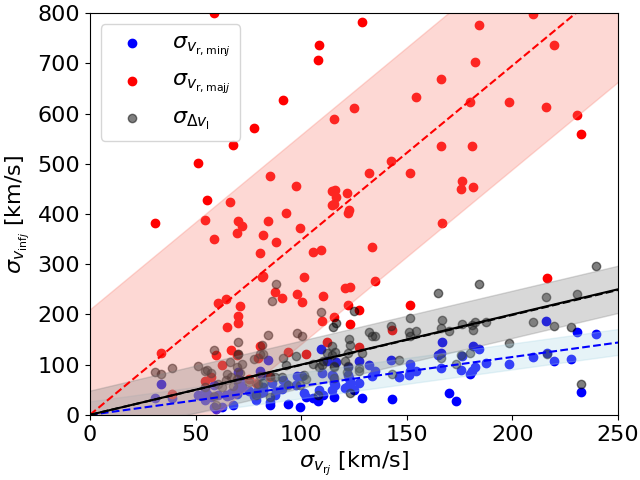}
\caption{\fixme{Velocity dispersions of bound subhalos for halos containing at least 5 bound subhalos $j=1,...,109$, calculated based on the minor and major infall models, as well as the infall velocity approximated by Eq.~\eqref{eq:theta0}.}}
\label{fig:vel_disp}
\end{figure}
While the infall model velocities themselves cannot constrain the radial velocity in a robust way, Fig.~\ref{fig:vel_disp} shows that the velocity dispersions are more suitable to constrain the true velocity dispersion. 
The plot shows the velocity dispersions of the bound subhalos for all 109 halos containing at least 5 bound subhalos. 
With only a few exceptions, the minor infall model systematically underestimates the true velocity dispersion, and the major infall model, even with a larger spread, systematically overestimates it. 
The infall velocity of Eq.~\eqref{eq:theta0} seems to yield the closest match to the true velocity dispersion. 
However, this result may not hold for cosmic structures at closer distances to the observer with larger $\theta$. 
Fitting a line with a 1-$\sigma$ confidence bound to all three velocity dispersion estimates, we obtain for the minor infall model $m_\sigma = 0.58$, $b_\sigma=25.78$~km/s, for the major infall model $m_\sigma = 3.48$, $b_\sigma=208.36$~km/s, and for the velocity difference model of Eq.~\eqref{eq:theta0} $m_\sigma = 0.99$, $b_\sigma=47.06$~km/s. 
In order to discard the quasi numerically unstable major infall velocities for which $|\boldsymbol{r}_{\mathrm{cm}j}| \approx 0$, we cut the absolute value at 3000~km/s, omitting 2\%, 31 out of 1476, bound subhalos. 
Hence, the plotted major-infall velocity dispersions are marginally lower compared to including these outliers.
For a set of observed galaxy groups in the same mass and distance range as our simulated selection, the parameters from the line fits could also be used to calibrate the velocity dispersions as obtained from the data in order to alleviate the model-based biases (see Sect.~\ref{sec:application} for an example).
}

\fixme{
Given these findings, any quantity which is dependent on the velocity dispersion can have a robust upper and lower bound based on the velocity dispersions of the minor and major infall model, or even a strong estimate based on the velocity dispersion of Eq.~\eqref{eq:theta0}, if the cosmic structure is far away enough from the observer.
One example for such a quantity is the virial mass. 
In contrast to this, any quantity which is dependent on the infall velocity itself may not have a robust upper and lower bound when the infall models are employed. 
One example is the $m_0$- and $r_0$-estimates based on Hubble-flow fits, as performed in \cite{bib:Kim2020}. 
It still remains an open question in how far a Hubble-flow fit to the minor and major infall models robustly yields upper and lower bounds on $m_0$ and $r_0$. 
Since this may also depend on the specific version of the Hubble-flow fitting function, this analysis is beyond the scope of this work. 
}

\fixme{
\begin{figure*}
\centering
\includegraphics[width=0.32\linewidth]{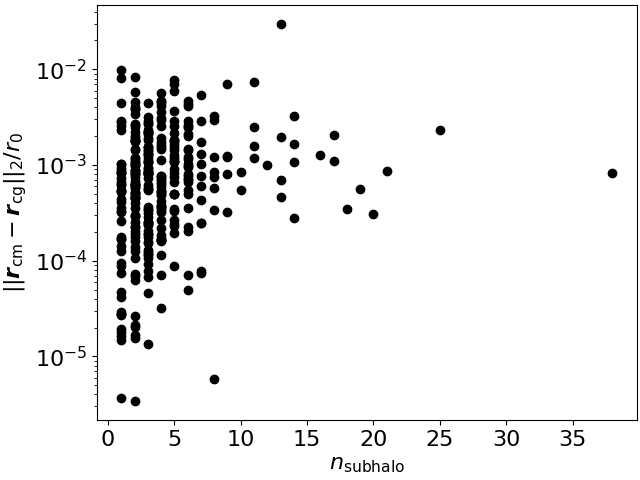} \hfill
\includegraphics[width=0.32\linewidth]{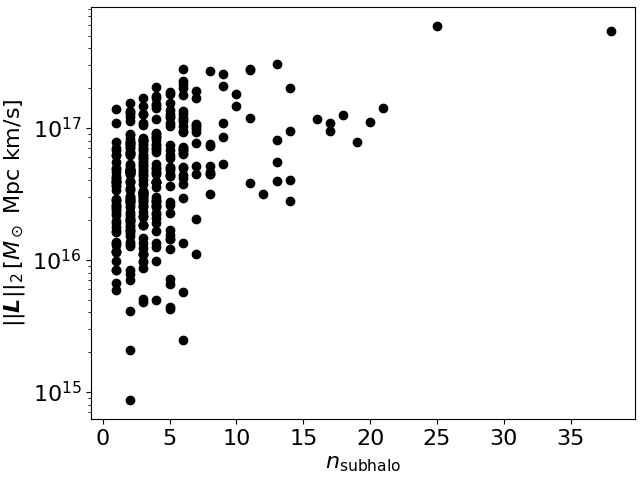} \hfill
\includegraphics[width=0.32\linewidth]{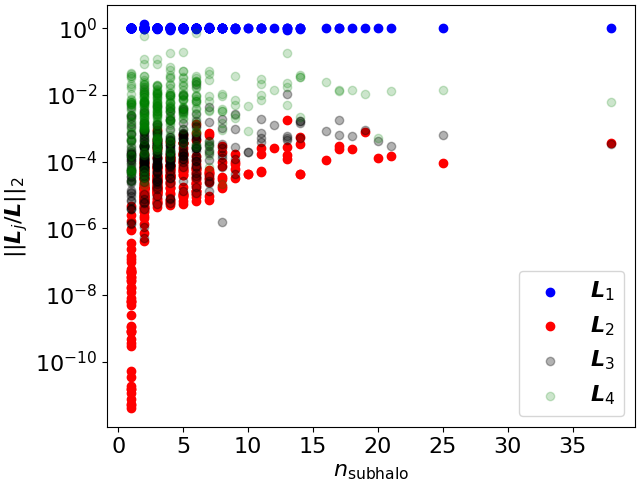}
\caption{\fixme{Left: Difference between the centre of mass of the parent halo and the centre of mass of the bound subhalos calculated as the mass-weighted sum of all their centre-of-mass positions versus the number of bound subhalos. The difference in the centre-of-mass positions is scaled to the zero-velocity radius of the halo for comparison. Centre: Total angular momentum as given by Eq.~\eqref{eq:L4} versus the number of bound subhalos. Right: Individual components of the angular momentum given by Eqs.~\eqref{eq:L_first}-\eqref{eq:L_fourth} with respect to the total one versus the number of bound subhalos.}}
\label{fig:L_tot}
\end{figure*}
At last, we investigate the assumptions made in Sect.~\ref{sec:galaxy_clusters} that motivate the major infall model. 
The Illustris-3 simulation has too low a resolution that we could test the impact of offsets between dark and luminous matter, i.e. whether the centre of light of the observed galaxies coincides with their centre of their mass and then infers the centre of mass of the galaxy cluster correctly.
So we can only analyse the ideal case, assuming we knew the centre of mass of the member galaxies of a cluster.
Fig.~\ref{fig:L_tot} (left) shows that the offset between the centre of mass of a halo and the centre of mass position of all its bound subhalos is very small compared to the halo extent, $r_0$. 
Thus, $\boldsymbol{r}_\mathrm{cm} = \boldsymbol{r}_\mathrm{cg}$ can be assumed for our halo selection.
}

\fixme{
Next, we determine the total angular momentum of a halo from all bound subhalos as defined by Eq.~\eqref{eq:L4} and plot its value against the number of bound subhalos in Fig.~\ref{fig:L_tot} (centre). 
Obviously, the angular momenta of the halos are far from vanishing. 
To sort the four individual parts of $\boldsymbol{L}$ as defined by Eqs.~\eqref{eq:L_first}-\eqref{eq:L_fourth}, we scale each of them to the total angular momentum and plot them in Fig.~\ref{fig:L_tot} (right). 
As can be read off Fig.~\ref{fig:L_tot} (right), the largest contribution to $\boldsymbol{L}$ comes from the angular momentum of the halo with respect to the observer. 
Hence, a vanishing overall angular momentum may be applicable to a large ensemble of halos that sample the volume around the observer well. 
However, for individual galaxy groups and clusters, this assumption does not hold.
Only the assumption that the angular momentum within the cosmic structure $\boldsymbol{L}_2$ is negligible compared to $\boldsymbol{L}_1$ is true to a good approximation for our selected halos. 
More interestingly, the contributions of $\boldsymbol{L}_3$ and $\boldsymbol{L}_4$ are generally larger then the one of $\boldsymbol{L}_2$, even for the small offsets between the halo centre of mass and the centre of mass inferred from its bound subhalos.
So we can conclude that the major infall model might be motivated by the derivations in Sect.~\ref{sec:galaxy_clusters}, yet, the assumptions made to arrive at Eqs.~\eqref{eq:v_maj_1} and \eqref{eq:v_maj_2} do not hold. 
Nevertheless, as we showed above, it is still a useful model to obtain an upper limit on the true radial velocity dispersion.
}

\section{\fixme{Application to the M81-group}}
\label{sec:application}

\fixme{
After investigating the validity and limits of the infall-model assumptions with simulations, we now apply the infall models to observations of the M81-group, whose Hubble flow was analysed in \cite{bib:Karachentsev2006}.
We will employ the recent data of M81 member galaxies as detailed in \cite{bib:Mueller2024}.
While there are more member galaxies known by now, we restrict our sample to the 21 of \cite{bib:Mueller2024}, as listed in Table~\ref{tab:M81-group}, because they base their collection on a comparably homogeneous data set to avoid self-inconsistencies due to potential biases caused by using various observational methods together.
As \cite{bib:Mueller2024} did not list the line-of-sight velocities and distances of M81 and M82, we added them from the same data base as used by \cite{bib:Mueller2024}, \cite{bib:Karachentsev2013}.
}

\fixme{Before we apply the infall models to the M81-group observables, we re-use the Illustris-3 simulation to select halos that have similar properties as those inferred from observations of the M81-group.
This will help to understand the large spread of the major-infall velocity dispersion in the Hubble flow around the M81-group that \cite{bib:Karachentsev2006} discovered.
It will also give an insight into the behaviour of the infall models at local-universe distances and we can compare the simulated velocity dispersion values with those from the observations. 
}

\subsection{\fixme{M81-group-like simulated structures}}
\label{sec:M81_simulated}

\fixme{
Similarly as in Sect.~\ref{sec:simulation}, we select halos and subhalos from Illustris-3 containing at least 30 and 20 particles, respectively. 
Based on the estimates of $r_0 \approx 1$~Mpc for the M81-group in \cite{bib:Karachentsev2006} and given that its environment has further infalling structures at a distance of about 1~Mpc \citep{bib:Chiboucas2013}, we select halos that do not have any other halo within $2\, r_0$ from their centre of mass. 
In addition, we require that the total halo mass $m_\mathrm{halo} \in \left[0.5, 5.0 \right] \times 10^{12} M_\odot$, based on the Hubble-flow estimate of \cite{bib:Karachentsev2006} that $m_\mathrm{M81} \approx 10^{12}~M_\odot$. 
The halo should also contain at least six subhalos, as the data by \cite{bib:Mueller2024} contains 11 galaxies inside what is assumed to be the second turn-around radius of 230~kpc and we allow for some fluctuations.
Since the M81-group has three strongly interacting galaxies in its centre, M81, M82, and NGC~3077 and seems to be a merging group, we allow a maximum offset between the centre of mass of the halo and the position of its most bound particle to be 100~kpc because this is the range of relative distances of the three interacting galaxies (see also Table~\ref{tab:M81-group}).
The position of the centre of mass of the halo is used in the following as the distance to the halo. 
For the subhalos, we do not set any limits on their relaxation to account for their potential interactions and employ the position of their most bound particle as their position.
}

\fixme{
\begin{figure*}
\centering
\includegraphics[width=0.33\linewidth]{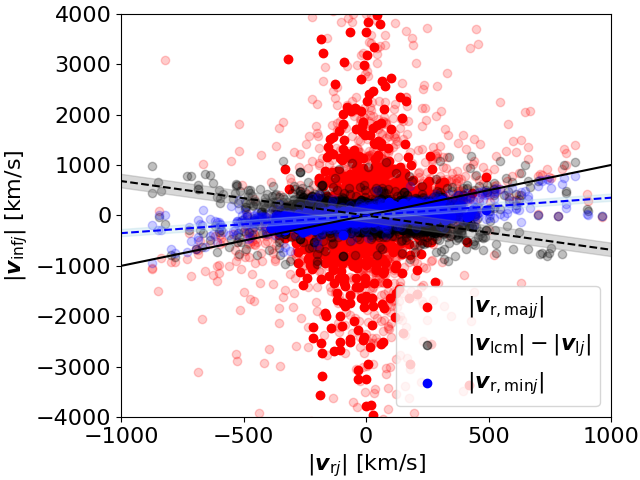} \hfill
\includegraphics[width=0.33\linewidth]{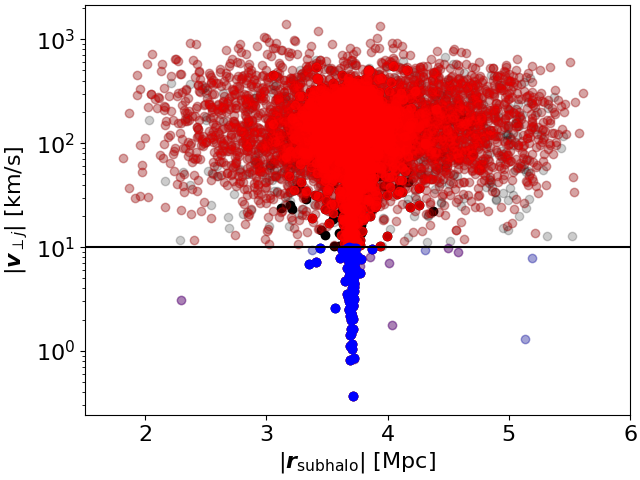} \hfill
\includegraphics[width=0.33\linewidth]{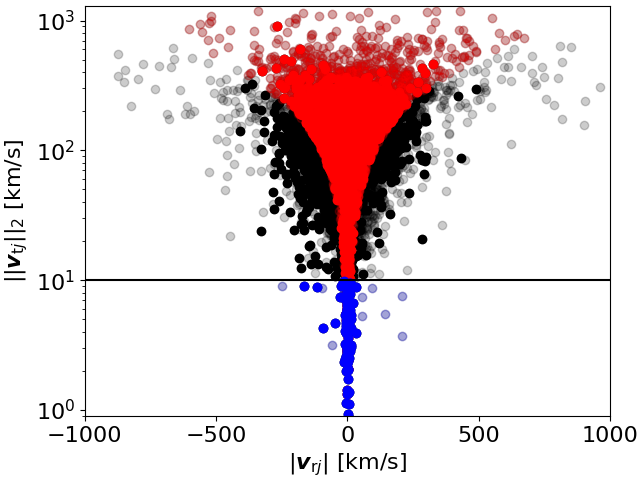}
\caption{\fixme{Same as Fig.~\ref{fig:vinf_vrad} but for the 281 M81-group-like halos. From all 5938 subhalos out to $1.5\, r_0$, 125 have a perpendicular velocity compatible with zero, 5427 have one that is larger than their line-of-sight velocity. For the bound subhalos, the numbers are 114 and 1925 out of 2099, respectively. 132 out of 5938 subhalos have a tangential velocity compatible with zero, for 5190, the tangential velocity exceeds the absolute value of the radial one. For the bound halos, the numbers are 118 and 1870, respectively.}}
\label{fig:M81_sim}
\end{figure*}
After selecting all halos and bound subhalos fulfilling these criteria, we also collect all subhalos around the halo out to 1.5~$r_0$ for the Hubble-flow analysis. 
Next, we set the observers' position with respect to each halo such that their distance to the centre of mass of the halo is 3.7~Mpc and their relative velocity is -38~km/s, which corresponds to the velocity by which M81 is approaching us as observers. 
}

\fixme{
Under these conditions, we obtain 281 halos to analyse. 
They contain a total 2099 bound subhalos and 5938 out to $1.5\, r_0$ with $n_\mathrm{subhalo} \in \left[6, 14\right]$ bound subhalos per halo and $n_\mathrm{subhalo} \in \left[7, 127\right]$ out into the Hubble flow. 
Fig.~\ref{fig:M81_sim} shows the same plots as in Fig.~\ref{fig:vinf_vrad} for the new halo selection. 
Compared to Fig.~\ref{fig:vinf_vrad}, the peculiar velocities of the galaxies play a larger role here, as they cause larger deviations between the infall models and the true radial velocity (left).
The perpendicular velocities exceed the line-of-sight velocities for 91\% of all subhalos (92\% of the bound ones) and are only compatible with zero in 2\% of all cases (5\% of the bound ones). 
Considering the tangential velocities, only 2\% of all subhalos (6\% of the bound ones) have such a component compatible with zero with a trend of jointly having a small radial velocity, similarly to the one observed in Fig.~\ref{fig:vinf_vrad}. 
However, 87\% of all tangential velocities exceed the absolute value of the radial component and 89\% out of all tangential velocities of bound subhalos.
On the whole, the results obtained for the infall models applied to more massive halos at larger distances are also found for less massive halos at smaller distances.
Even less subhalos fulfil the requirements of the infall models as stated in \cite{bib:Karachentsev2006} because the impact of the peculiar velocities on top of the cosmic expansion is stronger at smaller distances from the observer.
695 (33\%) of the bound subhalos have a radial velocity between the minor and major infall velocities, similarly, 1905 (32\%) of all subhalos into the Hubble flow fulfil this criterion. 
This is approximately the same ratio as for the far-away, more massive halos of Sect.~\ref{sec:simulation}.
}

\subsection{\fixme{Evaluation of observations and comparison}}
\label{sec:evaluation}

\begin{figure*}
\centering
\includegraphics[width=0.33\linewidth]{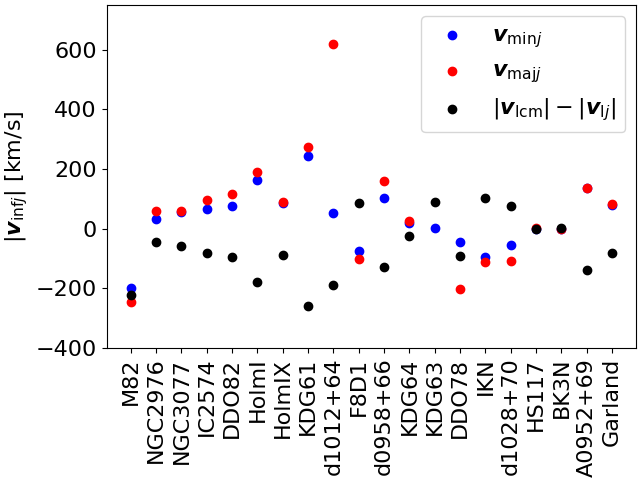} \hfill
\includegraphics[width=0.33\linewidth]{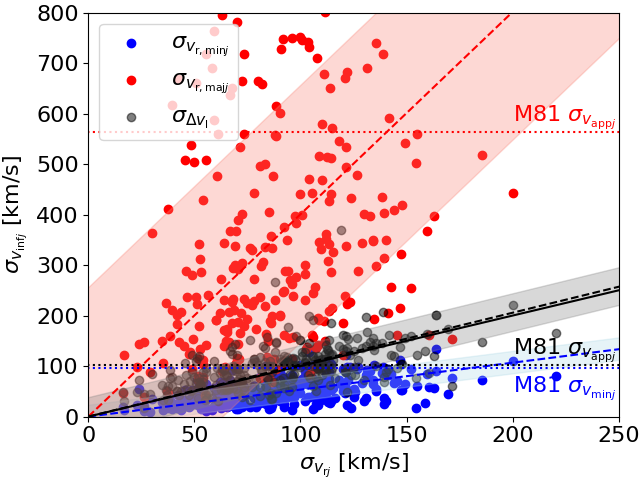} \hfill
\includegraphics[width=0.33\linewidth]{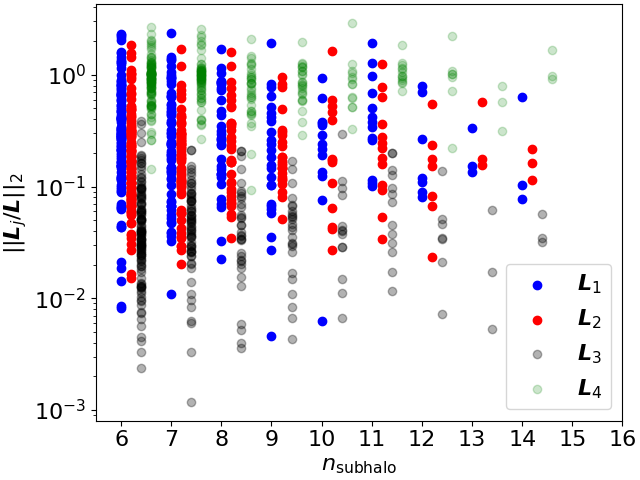}
\caption{\fixme{Left: Infall-model velocities defined in Eqs.~\eqref{eq:v_min_proj}, \eqref{eq:v_maj_1}, and \eqref{eq:theta0} for each member of the M81-group (see Table~\ref{tab:M81-group} for details on the data). Centre: Velocity dispersions of the infall models and the approximation of Eq.~\eqref{eq:theta0} of the bound subhalos in the simulation set up in Sect.~\ref{sec:M81_simulated} (same as Fig.~\ref{fig:vel_disp}) for $j=1,...,281$ halos. Dotted lines mark the velocity dispersions as determined for the M81-group members of Table~\ref{tab:M81-group}. Right: Ratios of absolute values of angular momenta as given by Eqs.~\eqref{eq:L_first}-\eqref{eq:L_fourth} with respect to the total angular momentum (same as Fig.~\ref{fig:L_tot}, right).}}
\label{fig:M81_eval}
\end{figure*}

Many data points listed in Table~\ref{tab:M81-group} lack proper measurement uncertainties.
Thus, we take the \revision{heliocentric velocity components without accounting for their uncertainties} to calculate the infall-model velocities and velocity dispersions of the member galaxies with respect to M81 as the centre for the infall. 
\revision{We then compare their results} to those obtained by the simulation set up in Sect.~\ref{sec:M81_simulated}\revision{, which amounts to the most direct comparison between observables and simulations possible}.
Fig.~\ref{fig:M81_eval} (left) shows the infall-model velocities and the velocity according to Eq.~\eqref{eq:theta0} for the individual member galaxies. 
As we can read off the plot, the spread in the infall velocities and even the large deviation between the infall velocity of Eq.~\eqref{eq:theta0} and the other infall models for many M81-group members resembles the findings of the simulations of Sect.~\ref{sec:M81_simulated}, even though the simulated structures may not fully represent the characteristics of the M81-group. 

In Fig.~\ref{fig:M81_eval} (centre), we apply the infall models to the bound subhalos of the M81-group-like simulated halos and plot their velocity dispersions.
Fitting lines with 1-$\sigma$ confidence bounds to the data points, we find for the minor infall model $m_\sigma = 0.53$, $b_\sigma=22.21$~km/s, for the approximation by Eq.~\eqref{eq:theta0} $m_\sigma = 1.03$, $b_\sigma=37.31$~km/s, and for the major infall model $m_\sigma = 4.01$, $b_\sigma=254.60$~km/s\footnote{As in Sect.~\ref{sec:simulation}, the fit excludes outliers with $||\boldsymbol{v}_{\mathrm{r,maj}}||_2>3000$~km/s. Yet, only 1\% (21) of all bound subhalos are affected by this cut.}.
Thus, we find similar results as for the halo selection of Sect.~\ref{sec:simulation}.

The dotted lines in Fig.~\ref{fig:M81_eval} (centre) are the velocity dispersions of the M81-group calculated for \revision{the heliocentric velocities, $v_\mathrm{hel}$, of} all members as listed in Table~\ref{tab:M81-group}. We obtain from the
\begin{align}
\text{minor infall}&: \sigma_{\mathrm{r,min}} = \revision{96}~\mbox{km}/\mbox{s} \;, \label{eq:M81_sigma_min}\\
\text{major infall}&: \sigma_{\mathrm{r,maj}} = \revision{564}~\mbox{km}/\mbox{s} \;, \label{eq:M81_sigma_maj}\\
\text{Eq.}~\eqref{eq:theta0}&: \sigma_{\mathrm{r,\Delta v_\mathrm{l}}} = \revision{102}~\mbox{km}/\mbox{s} \label{eq:M81_sigma_app}\;. 
\end{align}
\revision{
Often, heliocentric velocities are corrected for the motion of the sun with respect to a reference frame at larger scales. 
For the M81-group being a neighbour of the Local Group, \cite{bib:Karachentsev2013} also give all velocities of M81-members in the Local-Group-centroid frame, thus correcting for the solar motion within the Local Group, see Sect.~\ref{app:observational_data} for details.
Using these velocities given by $v_\mathrm{LG}$ in Table~\ref{tab:M81-group}, we obtain $\sigma_{\mathrm{r,min}} = 97~\mbox{km}/\mbox{s}$, $\sigma_{\mathrm{r,maj}} = 631~\mbox{km}/\mbox{s}$, and $\sigma_{\mathrm{r,\Delta v_\mathrm{l}}} = 102~\mbox{km}/\mbox{s}$.
Thus, given the size of the 1-$\sigma$ bounds on the simulated set of M81-like groups, the impact of this correction is likely to be absorbed in these bounds.}

\revision{Yet, it remains questionable, if this correction is reasonable to apply because it breaks the direct comparison to simulations, in which no such corrections are considered. 
The the Local-Group centroid is further moving with respect to larger scales, as is M81 and additional corrections may need to be considered. 
The infall models also assume that the unobserved velocity components are fine-tuned and transforming into one from the above mentioned reference frames doesn't necessarily match these conditions. Moreover, the heliocentric velocities correspond to us as observers and have the tightest error bars, while transformations into other reference frames, like the Local-Group centroid frame, accumulate additional uncertainties and model assumptions, such as the uncertainties from the proper motion of M31 (see~\cite{bib:derMarel2019} and~\cite{bib:Salomon2021}), the Local-Group mass ratio~\citep{bib:Karachentsev2008}, and that the dark matter is concentrated around the Milky Way and Andromeda~\citep{bib:Benisty2025b}. 
}

Assuming the M81-group to be a typical group obeying the average trends of the fitted lines, we can read off the most likely radial velocity for all three approximations \revision{of Eqs.~\eqref{eq:M81_sigma_min}-\eqref{eq:M81_sigma_app}} from Fig.~\ref{fig:M81_eval} (centre) as outlined to be a possible bias-correction method in Sect.~\ref{sec:simulation}:
\begin{align}
\text{minor infall}&: \sigma_{\mathrm{r}} = (\revision{180} \pm 42)~\mbox{km}/\mbox{s} \;, \\ 
\text{major infall}&: \sigma_{\mathrm{r}} = (\revision{142} \pm 64)~\mbox{km}/\mbox{s} \;, \\
\text{Eq.}~\eqref{eq:theta0}&: \sigma_{\mathrm{r}} = (\revision{99} \pm 36) ~\mbox{km}/\mbox{s} \;,
\end{align}
in which the error bounds are obtained from the 1-$\sigma$ confidence bounds of the linear fit. Due to the small angles, $\theta < 6$~deg, in the M81-group, we find Eq.~\eqref{eq:theta0} still to be a reliable approximation to the radial velocity dispersion. 

The latter approach is a simple application of Bayes' theorem 
\begin{equation}
P\left(\sigma_\mathrm{r} \; | \; \sigma_\mathrm{r,inf}= x \right) = \frac{P \left( \sigma_\mathrm{r,inf}= x \; | \; \sigma_\mathrm{r} \right) P\left(\sigma_\mathrm{r} \right)}{P\left( \sigma_\mathrm{r,inf}= x \right)} \;,
\label{eq:Bayes}
\end{equation}
in which we implicitly assumed a Gaussian likelihood for the velocity dispersions of the infall models, $P\left( \sigma_\mathrm{r,inf}= x \; | \; \sigma_\mathrm{r} \right)$ with $x$ given by Eqs.~\eqref{eq:M81_sigma_min} to \eqref{eq:M81_sigma_app}, and a uniform prior on the true radial velocity dispersion, $P\left(\sigma_\mathrm{r} \right)$.
Subsequently, we used the peak of the posterior distribution as the most likely radial velocity dispersion (coinciding with its Maximum A-Posteriori estimate) inferred from each infall model and the 1-$\sigma$ spread around it as the credible interval. 
Yet, as an Anderson-Darling test shows, these requirements are not well matched by the 281 simulated halos.  
To improve on the estimates, we create a numerical approximation of the posterior distribution. 
We keep the assumption of a uniform prior on $\sigma_\mathrm{r}$, but infer the likelihood function from the distribution of sample halos in Fig.~\ref{fig:M81_eval} (centre) for each infall model and the approximation of Eq.~\eqref{eq:theta0}.  
Then, we take the maximum A-Posteriori estimate again to infer the most likely value for the radial infall velocity. 
To implement this numerically, we partition the data into bins of \revision{25}~km/s width. Performing the Bayesian analysis in Eq.~\eqref{eq:Bayes}, we obtain from the
\begin{align}
\text{minor infall}&: \sigma_{\mathrm{r}} \approx \revision{150}~\mbox{km}/\mbox{s} \;, \\ 
\text{major infall}&: \sigma_{\mathrm{r}} = \revision{88}~\mbox{km}/\mbox{s} \;, \\
\text{Eq.}~\eqref{eq:theta0}&: \sigma_{\mathrm{r}} = 100~\mbox{km}/\mbox{s} \;.
\end{align}
Hence, even with this very sparse statistics, the results show a promising trend to use this approach as an estimator of the true radial velocity dispersion. 
With an increasing amount of simulated halos, it will also be possible to extend this simple model, for instance, by calculating the posterior under the constraint of the joint set of Eqs.~\eqref{eq:M81_sigma_min} to \eqref{eq:M81_sigma_app}.

While we do not investigate the velocity dispersion on top of the Hubble flow around the M81-group as \cite{bib:Karachentsev2006}, we still confirm the increase in the spread of the major-infall-model results compared to those of the minor-infall-model results for the group itself. 
Moreover, as we discovered in \cite{bib:Benisty2025}, we can confirm the increased spread of the infall velocities around the Hubble flow when analysing the outskirts of the Coma cluster, whose turnaround radius is approximately about 7~Mpc and whose mass is around $10^{15}~M_\odot$. 
Taken altogether, these results show that the definitions of the infall models as purely kinetic estimates and once as a symmetric projection (minor infall model), once as an asymmetric projection resulting in a velocity-distance ratio (major infall model) cause the effects that seemed surprising in \cite{bib:Karachentsev2006}. 

\fixme{
At last, we briefly comment on the assumption of spherical symmetry for the M81-group-like simulated halos. 
Fig.~\ref{fig:M81_eval} (right) shows the absolute values of the angular momenta compared to the total one, analogously to Fig.~\ref{fig:L_tot} (right). 
Since the halos are at closer distances to us, we find that the total angular momentum cannot be approximated as the angular momentum of the entire halo with respect to the observer anymore. 
The intrinsic angular momentum has now a comparably large value. 
Analogously to Fig.~\ref{fig:L_tot}, we additionally see that the offset between the centre of mass of the subhalos and the halo itself plays a significant role as well and cannot be neglected. 
We also observe a slightly decreasing trend of intrinsic angular momentum $||\boldsymbol{L}_2||_2$ with increasing number of subhalos. 
It may come from the sparser statistics, but it could also imply that the sampling comes more spherically symmetric with increasing number of samples, similarly as observed for the halo selection of Sect.~\ref{sec:simulation}. 
To corroborate this hypothesis, simulations with a better resolution and larger number of subhalos need to be evaluated.
}

\section{Conclusions}
\label{sec:conclusions}

We described an arbitrary binary motion on a general background including the velocity components perpendicular to the observer's lines of sight and the tangential velocity. 
In doing so, we generalised the radial infall models of \cite{bib:Karachentsev2006} that predict the relative velocity between pairs of galaxies or a galaxy falling towards the centre of mass of a group or cluster using the observed line-of-sight velocity components only. 

Both models are just different approximations to the general description when different information is available, as summarised in Fig.~\ref{fig:summary}:
For galaxies having small angular separations, all infall models agree that their relative radial velocity is the difference of their line-of-sight velocity components, Eq.~\eqref{eq:theta0}.
The M81-M82 binary system from \cite{bib:Karachentsev2006}, which will be revisited in \cite{bib:Benisty2025a}, is such an example. 
For arbitrary angles, we derived how the original models under- or overestimate the true radial velocity depending on the distance ratio, Eq.~\eqref{eq:conditions}. 
This extends the findings of \cite{bib:Kim2020} why minor and major infall models yield different $H_0$-values and that the true radial velocity need not be between the ones inferred from both models.  
In Eq.~\eqref{eq:equality}, we also show that the difference between the two infall models can be larger or smaller than zero depending on the difference between the line-of-sight velocities.

\fixme{
Applying the infall models and their small-angle approximation given by Eq.~\eqref{eq:theta0} to 344 isolated, relaxed halos of the Illustris-3 simulation in a mass range of $m \in \left[1.4, 16.4 \right] \times 10^{12}~M_\odot$ and at distances from an observer in the range of $d \in \left[18,162 \right]$~Mpc, we found that the following assumptions made in \cite{bib:Karachentsev2006}, \cite{bib:Karachentsev2010}, and \cite{bib:Kim2020} do not hold: 
even for small separation angles $\theta < 10$~deg of subhalos with respect to the centre of their parent halo, neither the velocity components perpendicular to the observers' line of sight, nor the tangential velocity components vanish. 
Fig.~\ref{fig:vinf_vrad} documents all the deviations from the true radial infall velocity and the expected vanishing velocity components.
So we concluded that it is not sufficient to select galaxies at large distances from an observer to make perpendicular velocity components vanish, nor is it sufficient to select galaxies in front and behind their cluster centre along the observers' line of sight to reduce the impact of the tangential velocity components.
Moreover, as Fig.~\ref{fig:Kim_discussion} revealed, it is not true that the true infall velocity always lies between the major and minor infall velocities. 
Overall, this only applied to 34\% of all subhalos bound to the 344 halos and out to 1.5 times the zero-velocity radius into their Hubble flow region. 
In addition to that, we discovered that the possible motivation of the major infall model based on a vanishing total angular momentum of a cosmic structure is invalid as well. 
The total angular momenta of the halos in our selection are non-zero and their largest contribution originates from the angular momentum of the halo with respect to the observer. 
While the angular momentum within the halos is compatible with zero compared to the angular momentum of the entire structure with respect to the observer, as shown in Fig.~\ref{fig:L_tot} (right), the contributions of the angular momenta due to an offset between the centre of mass of the halo and the centre of mass of all its bound subhalos were non-negligible. 
Thus, even in the ideal case, when we know the centre of mass and all masses of the subhalos, the additional constituents of the halo yield a contribution to the angular momentum that may be relevant. 
}

\fixme{
More positively, we also discovered that the velocity dispersions inferred from the infall models yield robust upper and lower bounds on the true velocity dispersion, as depicted in Fig.~\ref{fig:vel_disp}.
This allows us to constrain the true velocity dispersion and all quantities that depend on it much more robustly than any quantities depending on the infall velocities directly. 
For the large halo distances to the observer considered in our selection $d>18$~Mpc, the velocity dispersion based on the infall velocity in Eq.~\eqref{eq:theta0} even falls tightly onto the true velocity dispersion. 
So this may even be a good direct estimator instead of only upper and lower bounds for cosmic structures at large distances.
}

Applying the infall models to the M81-group also analysed in \cite{bib:Karachentsev2006}, we arrived at similar results for a newer dataset as detailed in \cite{bib:Mueller2024}. 
Yet, we could explain the large spread between the minor and major infall models as supported by a second selection of halos from the Illustris-3 simulation to imitate M81-like groups. 
For the latter simulations, we corroborated our results and thereby also showed that the infall models are purely based on kinematics and are thus independent of the masses involved in the models. 
Since the small-angle approximation also holds for the M81-group and 99\% of all selected halos being M81-group-like have $\theta < 10$~deg, the results of our latter simulation are also similar in that respect. 
The main difference between the two selected halo sets is that the perpendicular velocity components as well as the intrinsic angular momentum of the structure increase in their importance at closer distances.
From the observations of the M81-group, we arrive at the infall model approximations to the radial velocities as $\sigma_{\mathrm{r,min}}\approx \revision{96}$~km/s, $\sigma_{\mathrm{r,maj}}\approx \revision{564}$~km/s, $\sigma_{\mathrm{r,\Delta v}}\approx \revision{102}$~km/s. 
As the M81-group falls under the small angle approximation, the latter estimate is most likely to closest to the true radial velocity dispersion.
A similar result is also obtained with the more elaborate Bayesian inference of the radial velocity dispersion from the infall model approximations (see Sect.~\ref{sec:evaluation}), which, however, requires a larger amount of simulated M81-group-like halos in order to cover the feature space of all possible velocity dispersions in a more representative way. 

\revision{
One important aspect of observational data is the choice of reference frame. 
Corrections to account for the Earth's motion around the sun are necessary, as they directly affect all lines of sight to all structures.
The heliocentric velocities are measured to high precision, such that they are considered as observables seen from our cosmic position. 
Further corrections for the solar motion in the Milky Way, the Local-Group centroid or beyond up to the Cosmic Microwave Background vary in their impact along different lines-of-sight and are subject to much larger uncertainties, see \cite{bib:Aluri2023} for a recent overview of all relative motions and their tensions in our current concordance cosmology. 
Moreover, the infall models require fine-tuned assumptions about the velocity components to be accurate and transforming into one of these reference frames doesn't necessarily match these assumptions.
While studies of bulk flows or inferences of cosmological parameters in the local universe are best performed in the fundamental rest frame of matter (see \cite{bib:Ellis1985} and \cite{bib:Maartens2024} for more details), they often rely on a sufficiently large local volume containing so many cosmic structures that the approximation of a pressureless dust density is valid.
In contrast, our analysis of the M81-group is different, as we are studying an \emph{individual} cosmic structure. 
Thus, it is questionable whether any corrections for relative motions with respect to standard rest frames should be applied, as the transformations accumulate further uncertainties without making the infall models more accurate.
For the corrections of the solar motion within the Local Group, we found that they are so small that they are likely to be absorbed in measurement uncertainties anyway.
}

\fixme{
In summary, the infall models cannot constrain the true radial infall velocity as originally expected due to the surprisingly large contributions of the perpendicular and tangential velocity components. 
However, the velocity dispersion calculated from the infall model velocities is a more robust quantity to constrain the true velocity dispersion and quantities depending on it. 
This is even true, if the structure under analysis is not spherically symmetric, or is only sparsely sampled, as shown by the simulated halos with only a few number of subhalos.
Potentially, further summary-statistics based on a set of infall velocities could also be made robust bounds. 
One example is the Hubble flow fit to an ensemble of infall velocities. 
}

\fixme{
In how far the infall models can yield robust upper and lower bounds for binaries thus remains an open question. 
One possibility is to investigate if observations of tangential velocity components projected on the sky can alleviate the deviations or whether it is necessary to include satellite galaxies to the pair. 
An ideal system to study is the Local Group, for which respective observables are available \citep{bib:Benisty2022}.
}

\fixme{
For galaxy clusters of masses above $10^{14}M_\odot$ with velocity dispersions of 1000~km/s, the difference in the velocity dispersions of the major and minor infall models is assumed to be larger than for our halo selection with the true velocity dispersion being closer to the one of the minor infall model. 
Hence, in contrast to the original motivation for the major infall model, the minor infall model may yield more accurate velocity dispersions and dependent quantities than the major infall model for galaxy clusters.
A further simulation of high-redshift clusters will also reveal if the velocity dispersion given by Eq.~\eqref{eq:theta0} actually yields the tightest constraint on the true velocity dispersion. 
If true, virial mass estimates of high-redshift clusters based on the difference of line-of-sight velocity components will be confirmed to be the most accurate estimate with the tightest confidence bounds.
More generally, as corroborated by our study of the M81-group, any structure for which the small-angle approximation holds can employ this result. 
For all other structures, the major and minor infall models provide the most robust upper and lower bounds to the radial velocity dispersion. 
} 

 \begin{figure}
     \centering
        \begin{tikzpicture}[auto,
                       > = Stealth, 
           node distance = 15mm and 34mm, 
              box/.style = {draw=black, thick,
                            minimum height=11mm, text width=22mm, 
                            align=center},
       every edge/.style = {draw, <->, thick},
every edge quotes/.style = {font=\footnotesize, align=center, inner sep=2pt}
                            ]
    \node (n11) [box]               {group motion \\ $n > 2$ \\ Eqs.~\eqref{eq:P} \eqref{eq:L4}};
    \node (n12) [box, right=of n11] {major infall model \\ Eqs.~\eqref{eq:v_maj_1} \eqref{eq:v_maj_2}};
    \node (n21) [box, above=of n11] {binary motion \\ $n=2$ \\ Eqs.~\eqref{eq:v_r_proj} \eqref{eq:v_t_proj}};
    \node (n22) [box, above=of n12] {minor infall model \\ Eq.~\eqref{eq:v_min_proj}};
\draw   (n11) edge [<-, "increase \\ samples"]     (n21)
        (n21) edge [->,"$\theta=0$ or Eq.~\eqref{eq:conditions}"]         (n22)
        (n22) edge ["$\theta=0:$ equal \\ $\theta \ne 0$: Eq.~\eqref{eq:equality}"]    (n12)
        (n12) edge [<-, above, sloped, pos=0.5, "$\theta = 0$ or Eq.~\eqref{eq:v_maj_cond}"']  (n21)
        (n11) edge [->,above,"$\boldsymbol{r}_\mathrm{cm} = \boldsymbol{r}_\mathrm{cg}$ and"]      (n12)
        (n11) edge [->,below, "spher. symm. in Eq.~\eqref{eq:L_maj}"] (n12);
        \end{tikzpicture}
    \caption{Summary of all results: conditions to apply the infall models, their formulae, relations, and accuracy.}
   \label{fig:summary}
    \end{figure}

\begin{acknowledgements}
\fixme{We cordially thank the anonymous referee for suggesting valuable improvements, Rainer Weinberger, Dylan Nelson and Volker Springel for their further comments and discussion on the TNG simulations \revision{and Roy Maartens for his help on the rest-frame issue}.} DB is supported by a Minerva Fellowship of the Minerva Stiftung Gesellschaft f\"ur die Forschung mbH.
\end{acknowledgements}

%-------------------------------------------------------------------
\bibliographystyle{aa}
\bibliography{ref}

\begin{appendix}
\onecolumn
\section{Observational data set}
\label{app:observational_data}
\revision{
The data shown in Table~\ref{tab:M81-group} is taken from \cite{bib:Mueller2024} except for the entries of M81 and M82 which are taken from \cite{bib:Karachentsev2013}.
As stated in the latter, the measured heliocentric radial velocities are corrected for the motion of the sun within the Local-Group centroid frame as
\begin{align}
v_\mathrm{LG} = v_\mathrm{hel} &+ 316~\tfrac{\mbox{km}}{\mbox{s}} \, \Big( \sin (b) \sin(-4^\circ) + \cos(b) \cos(-4^\circ) \cos(l-93^\circ) \Big) \;,  
\end{align}
in which $b$ and $l$ denote the coordinates of the galaxy in the Galactic coordinate system (following the IAU’s 1958 definition).
The apex velocity of $316$~km/s and the apex coordinates of $(93^\circ, -4^\circ)$ in Galactic longitude and latitude are taken from \cite{bib:Karachentsev1996}.
As stated in this work, uncertainties in the coordinates amount to $2$~deg and to 5~km/s in the apex velocity. 
They need to be taken into account to obtain confidence bounds on $v_\mathrm{LG}$.
Consequently, uncertainties after applying the correction are larger than those in the heliocentric frame.
No uncertainties are listed for $v_\mathrm{LG}$ in the data base of \cite{bib:Karachentsev2013}. 
As our proof-of-principle analysis focusses on accuracy and not precision, $v_\mathrm{LG}$ are listed without uncertainties, but the latter should be included in a full analysis of the data.
}

\begin{table*}[h!]
\caption{M81-group members.}
\label{tab:M81-group}
\centering
\begin{tabular}{cccccccc}
\hline\hline  
Name & RA [deg] & Dec [deg] & $\mathrm{r}$ [Mpc] & $M_r$ [mag] & $v_{\text{hel}}$ [km/s] & \revision{$v_{\text{LG}}$ [km/s]} & bound\\
\hline
  M81 & 148.88821 & 69.06528 & $3.7 \pm 0.0$ & -20.6 & $-38.0 \pm 21.0$ & \revision{104} & $\checkmark$ \\
  M82 & 148.96846 & 69.67970 & $3.61 \pm 0.00$ & -19.9 & $183.0 \pm 0.0$ & \revision{328} & $\checkmark$\\
  NGC2976 & 146.81500 & 67.91361 & $3.7 \pm 0.1$ & -18.0 & $6.0 \pm 4.0$ & \revision{142} & $\checkmark$\\
  NGC3077 & 150.83750 & 68.73389 & $3.8 \pm 0.1$ & -17.8 & $19.0 \pm 4.0$ & \revision{159} & $\checkmark$\\
  IC2574 & 157.09333 & 68.41611 & $3.9 \pm 0.0$ & -17.7 & $43.0 \pm 4.0$ & \revision{183} & \\
  DDO82 & 157.64583 & 70.61944 & $3.9 \pm 0.0$ & -15.1 & $56.0 \pm 3.0$ & \revision{207} & \\
  HolmI & 145.13458 & 71.18639 & $4.0 \pm 0.1$ & -14.6 & $139.4 \pm 0.1$ & \revision{291} & \\
  HolmIX & 149.38500 & 69.04306 & $3.8 \pm 0.1$ & -13.9 & $50.0 \pm 4.0$ & \revision{192} & $\checkmark$\\
  KDG61 & 149.26125 & 68.59167 & $3.7 \pm 0.0$ & -13.3 & $221.0 \pm 3.0$ & \revision{360} & $\checkmark$\\
  d1012+64 & 153.20171 & 64.10722 & $3.7 \pm 0.1$ & -13.3 & $150.0 \pm 50.0$ & \revision{267} & \\
  F8D1 & 146.19625 & 67.43861 & $3.8 \pm 0.1$ & -13.2 & $-125.0 \pm 130.0$ & \revision{8} & $\checkmark$\\
  d0958+66 & 149.70308 & 66.84917 & $3.8 \pm 0.1$ & -13.2 & $90.0 \pm 50.0$ & \revision{221} & \\
  KDG64 & 151.75792 & 67.82750 & $3.8 \pm 0.0$ & -13.2 & $-15.0 \pm 13.0$ & \revision{121} & $\checkmark$\\
  KDG63 & 151.28042 & 66.55500 & $3.6 \pm 0.0$ & -13.0 & $-129.0 \pm 0.3$ & \revision{0} & $\checkmark$\\
  DDO78 & 156.61625 & 67.65667 & $3.5 \pm 0.0$ & -12.8 & $55.0 \pm 10.0$ & \revision{191} & $\checkmark$\\
  IKN & 152.02458 & 68.39917 & $3.8 \pm 0.0$ & -12.4 & $-140.0 \pm 64.0$ & \revision{-1} & $\checkmark$\\
  d1028+70 & 157.16658 & 70.23333 & $3.8 \pm 0.1$ & -12.4 & $-114.0 \pm 50.0$ & \revision{35} & \\
  HS117 & 155.35500 & 71.11611 & $4.0 \pm 0.1$ & -12.1 & $-37.0 \pm 0.0$ & \revision{116} & \\
  BK3N & 148.45208 & 68.96917 & $4.2 \pm 0.3$ & -10.3 & $-40.0 \pm 0.0$ & \revision{101} & \\
  A0952+69 & 149.37083 & 69.27222 & $3.9 \pm 0.3$ &  & $99.0 \pm 0.0$ & \revision{242} & \\
  Garland & 150.92500 & 68.69333 & $3.8 \pm 0.4$ &  & $43.0 \pm 17.0$ & \revision{183} & $\checkmark$\\
\hline
\end{tabular}
\tablefoot{
Data taken from \cite{bib:Mueller2024} with their name, coordinates on the sky (J2000), distance from us as observers, magnitude in the $r$-band and heliocentric velocity. The missing $v_\mathrm{hel}$ and $r$ for M81 and M82 are filled in using the database of \cite{bib:Karachentsev2013}, as done by \cite{bib:Mueller2024}. The last column indicates whether the galaxy is within a three-dimensional distance of 230~kpc around M81 and thus within the radius of the second-turnaround radius.
}
\end{table*}

\end{appendix}

\end{document}